# Non-Parametric and Semi-Parametric Asset Pricing

Péter Erdős
Department of Finance,
Budapest University of Technology and Economics
Budapest, Műegyetem rkp. 9., 1111. HUNGARY
erdos@finance.bme.hu
Phone: +36 1 463 1083, Fax +36 1 463 2745

Mihály Ormos*
Department of Finance,
Budapest University of Technology and Economics
Budapest, Műegyetem rkp. 9., 1111. HUNGARY
ormos@finance.bme.hu
Phone: +36 1 463 4220, Fax +36 1 463 2745

Dávid Zibriczky
Department of Computer Science and Information Theory,
Budapest University of Technology and Economics
Budapest, Műegyetem rkp. 9., 1111. HUNGARY
zibriczky@finance.bme.hu
Phone: +36 70 259 9901, Fax +36 1 463 2745

*corresponding author

# Non-Parametric and Semi-Parametric Asset Pricing

**Abstract**

We find that the CAPM fails to explain the small firm effect even if its non-parametric form is used which allows time-varying risk and non-linearity in the pricing function. Furthermore, the linearity of the CAPM can be rejected, thus the widely used risk and performance measures, the beta and the alpha, are biased and inconsistent. We deduce semi-parametric measures which are non-constant under extreme market conditions in a single factor setting; on the other hand, they are not significantly different from the linear estimates of the Fama-French three-factor model. If we extend the single factor model with the Fama-French factors, the simple linear model is able to explain the US stock returns correctly.

# 1. Introduction

The Capital Asset Pricing Model (CAPM, Sharpe, 1964; Lintner, 1965; Mossin, 1966) is the most broadly applied equilibrium model in financial literature and among practitioners. However, despite its success, there are many studies denying the validity of the CAPM. Among others, Banz (1981), Basu (1983), Bhandari (1988) and Fama and French (1995) provide evidence against the CAPM. Researchers try to override the problem of the CAPM by alternative models which have fallen in three classes: (1) Multifactor extensions of the linear model such as the Merton's ICAPM (1973) or Ross' (1976) APT; (2) conditional versions of the linear model allowing time-variation in the market risk; (3) non-linear asset pricing models. The Fama-French three-factor-model (Fama and French, 1996) extends the single factor model adding two risk factors, one for the small firm effect and one for the 'relative distress'. Carhart (1997) involves the momentum factor which is also priced besides the three Fama-French factors. Keim and Stambaugh (1986) and Breen et al. (1989) show that conditional betas are not constant. While Fama and French (1989), Chen (1991), and Ferson and Harvey (1991) prove that betas vary over the business cycles. Ferson (1989), Ferson and Harvey (1991, 1993), Ferson and Korajczyk (1995), and Jaganathan and Wang (1996), among others, provide further evidence that the market risks vary from time to time. Jagannathan and Wang (1996) formalize a conditional CAPM which exhibits some empirical evidence that betas are time-varying. However, they argue that firm size effect is not significant in this setting. Zhang (2006) finds that the conditional international CAPM with exchange risk provides the least pricing error. However, Stapleton and Subrahmanyam (1983) verify a linear relationship between risk and return in the case of the CAPM; there are many studies that contradict this result. Barone Adesi (1985) proposes a quadratic three-moment CAPM. Bansal and Viswanathan (1993) show that a non-linear, two-factor model, extending the market risk factor with the one-period yield in the next period, outperforms the CAPM. Bansal et al. (1993) shows that a non-linear arbitrage-pricing model is superior over the linear conditional, and the linear unconditional models, for pricing international equities, bonds and forward currency contracts. Chapman (1997) argues that non-linear pricing kernel in CCAPM performs better than the standard CAPM. Dittmar (2002) argues that a cubic pricing kernel induces much less pricing error than a linear one. Asgharian and Karlsson (2008) confirm the pricing ability of the non-linear model suggested by Dittmar (2002) on international equity, allowing time varying risk prices. Akdeniz et al. (2002) elaborate a new non-linear approach that allows time-varying betas and is called threshold CAPM. The model allows beta to change when the threshold variable hits a certain threshold level.

Linear asset pricing test is adequate only if the relationship between risk and return is indeed linear. If this assumption does not hold, then the estimated parameters by the Ordinary Least Squares (OLS) or by any other linear estimators are biased and inconsistent. In our study we pick the returns of 50-50 randomly chosen stocks from each size index of the S&P universe (from the S&P 500, S&P MidCap 400, S&P SmallCap 600) from the Center of Research in Stock Prices (CRSP) database for the period of 1999 to 2008.

First, we investigate whether or not a single factor model based on kernel regression can explain the US stock returns allowing time varying market risk and non-linear asset pricing. We estimate a time-series regression of the risk premium of a given security on the market risk premium which relationship is known as the characteristic line (CL). Based on the estimated non-parametric CLs we can test the linearity and the stability of betas as well. We find that the linear nullity of the CL, as a general rule for the whole market, can be rejected at any usual significance level. Our non-parametric asset pricing outperforms the linear one as it induces higher $R^2$. As the linearity of the CLs is rejected, we derive semi-parameters which override the problem of the linear regression and provide a mean average square error (MASE) optimal estimates. We show that the semi-parameric beta, which is the average slope coefficient of the non-parametric CL, is not constant when extreme market movements occur. On the other hand, the proposed semi-parametric analysis allows us not just to test the stability of the market risk, but that of the performance measurement known as the Jensen alpha. We show that portfolio managers can beat the market only when extreme market movements occur, since the abnormal return is constant and significantly not different from zero under normal market movements.

Second, we investigate the cross sections of the US stock returns and regress the expected asset return on its market risk measured by the semi-parametric beta; this relationship is known as the security market line (SML). We estimate four cross sectional regressions; the linearity of the SMLs cannot be rejected. The slope of the SML for the S&P small companies is negative; that is, our results confirm the small company effect (see, e.g., Banz, 1981; Basu, 1983; Fama and French, 1995). It is well-documented in financial literature that small stocks have higher risk, thus they provide higher expected return. Contrary to Jagannathan and Wang (1996), our results confirm that firm size is a risk factor besides the market risk, even if we allow non-linear pricing and the market beta to vary. On the other hand, it is striking that linearity of the large cap CLs can be rejected at any usual significance level, while those of small and mid companies cannot be rejected at 95% significance level which means the small company effect does not explain the fallacy of linearity. Our results show that the most probable explanantions for non-linearity are the omitted risk factors and/or extreme market movements.

Finally, we estimate the non-parametric generalization of the Fama-French three-factor model as the single factor model does not hold. The linearity hypothesis of this model cannot be rejected at any level, which is also supported by the fact that the semi-parameters are not significantly different from the linear estimates. These results support that the linear three-factor setting can explain the US stock returns even not allowing time varying betas and non-linear asset pricing.

## 2. Methodology

A multivariate regression is a conditional expected value, in the most general form

$$E(Y \mid X) = m(X) \tag{1}$$

where $Y$ is the dependent variable and $X=(X_1, X_2, ..., X_d)$ is the regressing vector. Assuming linear relationship, simple linear estimators can be applied; however, if linearity does not hold, then the linear estimators induce biased and inconsistent parameter estimations. We need a distribution free, robust estimator for the tests sketched in the introduction section which induces precise estimations even in a non-linear environment. If we assume that Eq. (1) is linear, a multifactor model is

$$Y_{i,k} = \hat{\alpha}_k + \sum_{d=1}^{D} \hat{\beta}_k^d X_{i,k}^d + \hat{\varepsilon}_{i,k} \quad i=1,2,3,...,n; \quad k=1,2,3,...,N \tag{2}$$

where $Y_{i,k}$ is the risk premium of the $k$th security and $X_{i,k}^d$ $d=1,2,...,D$ are different regressors in period $i$, $\hat{\alpha}_k$ is the intercept, $\hat{\beta}_k = \left(\hat{\beta}_k^{d=1}, \hat{\beta}_k^{d=2}, ..., \hat{\beta}_k^{d=D}\right)$ is a vector of unknown parameters to be estimated and $\hat{\varepsilon}_{k,i}$ is the residual series of the regression.

We do not assume linear relationship between the variables; that is, linear regression is not a suitable method for estimating Eq. (1). Nadaraya (1964) and Watson (1964) derive a kernel based regression estimator, which can estimate Eq. (1) without assuming any specific form of the relationship between the variables. The Nadaraya-Watson estimator is

$$\hat{m}_H(x) = \frac{1}{n} \sum_{i=1}^{n} W_{Hi}(X_i) Y_i, \qquad (3)$$

where $X_i = \left(X_i^{d=1}, X_i^{d=2}, \ldots, X_i^{d=D}\right)$ is a matrix of explanatory variables and $W_{Hi}(x)$ is the Nadaraya-Watson weighting matrix; that is,

$$W_{Hi}(x) = \frac{K_H(x - X_i)}{\frac{1}{n} \sum_{j=1}^{n} K_H(x - X_j)}, \qquad (4)$$

where $K_H(u) = \prod_{d=1}^{D} k(u_d)$ is the multivariate kernel function and $H$ is a properly selected bandwidth matrix.

### 2.1. Choice of the kernel function and bandwidth selection

Härdle et al. (2004) show that the choice of the kernel function is only of secondary importance, so the focus is rather on the right choice of bandwidth. Most of the frequently used kernel functions (uniform, triangular, Epanechnikov, triweight, cosine, Gaussian) include an indicator, which is equal to one if the condition embedded in the function meets and is equal to zero otherwise. However, Härdle et al. (2004) argue that the Epanechnikov kernel has the fastest convergence, although estimating our results by different kernels, we find no significant difference[1].

Since we do not assume linearity and want to characterize the market risk by one single measure, we derive semi-parametric performance and risk measures ("alpha" and "beta") by derivative estimations, hence we need a kernel function which is differentiable at each point, thus we use the indicator free Gaussian kernel, which is in the form

$$K(u) = \frac{1}{\sqrt{2\pi}} exp\left(-\frac{1}{2}u^2\right). \tag{5}$$

---

[1] These results are not presented in this study but they are available upon request.

In the case of the single variate kernel regression, smoothing depends on bandwidth *h*, the kernel function becomes flatter as its value is growing, and thus the impact of closer values declines, while the impact of more distant values grows at each point of estimation. We select an optimal bandwidth which minimizes the Average Square Error (ASE) which is in the form

$$ASE(h) = ASE\{\hat{m}_h\} = \frac{1}{n}\sum_{i=1}^{n}\{\hat{m}_h(X_i) - m(X_i)\}^2, \tag{6}$$

where $\hat{m}_h(X_i)$ is the estimated form of Eq. (1) and $m(X_i)$ is the true value of the same. Since $m(X_i)$ is unknown we use the Cross Validation for the optimization in the form

$$G(h) = \frac{1}{n}\sum_{i=1}^{n}\{Y_i - \hat{m}_h(X_i)\}^2 \; \Xi\left(\frac{1}{n}W_{hi}(X_i)\right), \tag{7}$$

where $\Xi(h)$ is the penalizing function which grows as *h* declines; that is, it adjusts the error emerging from the naive approximation of $Y_i \sim m_h(x)$ (see Härdle et al., 2004). Let us assume the Generalized Cross-Validation penalizing function in the form

$$\Xi_{GCV}(u) = (1-u)^{-2}. \tag{8}$$

If we substitute Eq. (8) into Eq. (7), we obtain

$$CV(h) = \frac{1}{n}\sum_{i=1}^{n}\{Y_i - \hat{m}_h(X_i)\}^2 \left(1 - \frac{1}{n}W_{hi}(X_i)\right)^{-2}, \tag{9}$$

which is known as the Cross-Validation (CV) function. Härdle et al. (2004) show that the ASE is minimal when $CV(h)$ is minimal, thus the bandwidth based on Eq. (9) is optimal. We use the simplex search method for the minimizing problem (see Lagarias et al., 1998). The iteration process can be accelerated if we choose an initial value close to the optimum, for example, based on the Silverman's rule of thumb[2]. An adjusted version of the Silverman's (1986) rule of thumb is in the form

$$\hat{h}_{rot} = 1.06 \, min \left\{ \sqrt{\frac{1}{n-1} \sum_{i=1}^{n} \left( X_i - \overline{X} \right)^2 }, \frac{Q_3 - Q_1}{1.34} \right\} n^{-\frac{1}{5}}, \qquad (10)$$

where $Q_3$ and $Q_1$ are the third and the first quartile of $X_i$, respectively. The closer the distribution to the normal, the more accurately the rule works. Since our time series are not normally distributed, the Silverman selected bandwidth is not optimal; however, it is a suitable choice as an initial value for the minimizing algorithm which reduces the computational complexity efficiently.[3]

In the multivariate setting

$$H = \begin{bmatrix} h_1 & 0 & L & 0 \\ 0 & h_2 & L & 0 \\ M & M & O & 0 \\ 0 & 0 & L & h_d \end{bmatrix}$$

is a diagonal matrix which is optimal if $h_d$ d=1,2,…,D are optimally regressing $Y_{i,k}$ on each $X_{i,k}^d$ alone (see Härdle et al., 2004).

---

[2] It is necessary if time series are long, since the computational time of the algorithm minimizing the $CV(h)$ function is proportional to the fourth power of the number of observations.
[3] We reject the normality of all of the time series based on the Jarque-Bera test. This result is available upon request.

## 2.2. Goodness of fitting

To sustain comparability with the linear regression, we use $R^2$ as the measure of goodness of fitting. By definition, $R^2$ is

$$R^2 \equiv 1 - \frac{SSE}{SST}, \qquad (11)$$

where $SSE = \sum_{i=1}^{n}(Y_i - \hat{m}_H(X_i))^2$ and $SST = \sum_{i=1}^{n}(Y_i - \bar{Y})^2$.

This definition is equivalent to the $R^2$ used for linear regression; the difference is only that there is the kernel estimation, $\hat{m}_H(X_i)$ in $SSE$ instead of the parametric estimation. Since we calculate both measures in the same way, they are comparable[4].

## 2.3. Hypothesis testing and test statistic

We test whether or not the CL and the SML are linear by applying the nullhypothesis of linear regression against the kernel regression alternative. Assume the parametric model in the form

$$E(Y | X = x) = m_\theta(\bullet), \qquad (15)$$

where $\theta$ is a vector of parameters, thus the nullhypothesis is $H_0 : m(x) \equiv m_\theta(x)$, which is tested against the $H_1 : m(x) \neq m_\theta(x)$ alternative. The $\hat{\theta}$ vector is the estimation of $\theta$ which can be estimated by standard parametric regressions. $m(x)$ is unknown, thus we use $\hat{m}_H(x)$ to approximate it. If we cannot reject $H_o$, it means that the kernel regression does not differ

---
[4] In the case of linear regression it would be more adequate to use adjusted $R^2$ since we lose several degrees of freedom because of parameter estimation (in our case we lose two). We have to note that it has no significant impact because we use a relatively large sample.

significantly from the parametric one. The difference between the two estimations can be measured by

$$\sum_{i=1}^{n}\left\{\hat{m}(X_{i})-m_{\hat{\theta}}(X_{i})\right\}^{2}. \tag{16}$$

While $m_{\hat{\theta}}(\bullet)$ is asymptotically unbiased and the speed of convergence of the parameters is $\sqrt{n}$, the non-parametric estimation is biased because of smoothing and the speed of convergence is only $\sqrt{nh}$. Härdle and Mammen (1993) introduce an artificial bias into the parametric estimation to solve this problem. They use kernel weighted regression in the form

$$\hat{m}_{\hat{\theta}}(X_{i}) = \frac{\sum_{j=1}^{n} K_{H}(X_{i}-X_{j}) m_{\hat{\theta}}(X_{j})}{\sum_{j=1}^{n} K_{H}(X_{i}-X_{j})}, \tag{17}$$

instead of $m_{\hat{\theta}}(X_{i})$ and based on Eq. (16) they use

$$T = \sum_{i=1}^{n}\left\{\hat{m}_{H}(X_{i})-\hat{m}_{\hat{\theta}}(X_{i})\right\}^{2} \tag{18}$$

test statistic. The distribution of $T$ is unknown; however, it can be determined by the wild bootstrap approach (see Härdle and Mammen, 1993).

### 2.4. Risk and performance measurement, "beta" and "alpha" estimation

The relevant, non-diversifiable risk can be measured linearly by $\hat{\beta}$; that is, the slope coefficient of the CL. In the case of non-linearity, this beta estimation is biased, thus we approximate the market risk by a semi-parametric method.[5] Härdle et al. (2004) show that $\hat{\beta}^*(x) = (\hat{\beta}_0(x), \hat{\beta}_1(x), ..., \hat{\beta}_p(x))^T$ can be estimated by minimizing

$$\min_{\hat{\beta}_0, \hat{\beta}_1, ..., \hat{\beta}_p} \sum_{i=1}^{n} \left\{ Y_i - \hat{\beta}_0^* - \hat{\beta}_1^*(X_i - x) - ... - \hat{\beta}_p^*(X_i - x)^p \right\}^2 K_h(x - X_i). \tag{21}$$

$\hat{\beta}^*(x)$ can be estimated by the Weighted Least Squares (WLS) in the form

$$\hat{\beta}^*(x) = (X^T W X)^{-1} X^T W Y, \tag{22}$$

where the weights are defined in Eq. (4), $X = \begin{pmatrix} 1 & X_1 - x & (X_1 - x)^2 & L & (X_1 - x)^p \\ 1 & X_2 - x & (X_2 - x)^2 & L & (X_2 - x)^p \\ M & M & M & 0 & M \\ 1 & X_n - x & (X_n - x)^2 & L & (X_n - x)^p \end{pmatrix}$,

$p$ is the power of the regression, $Y = \begin{pmatrix} Y_1 \\ Y_2 \\ M \\ Y_n \end{pmatrix}$ and $W = \begin{pmatrix} K_h(x - X_1) & 0 & L & 0 \\ 0 & K_h(x - X_2) & L & 0 \\ M & M & 0 & M \\ 0 & 0 & L & K_h(x - X_n) \end{pmatrix}$.

The estimation defined in Eq. (22) is a local polynomial regression (see Härdle et al., 2004). The $\hat{\beta}^*(x)$ vector has as many elements as the power of the estimated equation, thus, for example, $\hat{\beta}_0^*(x)$ is the local constant estimation of the $\hat{m}_h(x)$ regression function, which is itself the Nadaraya-Watson kernel regression. $\hat{\beta}_1^*(x)$ approximates the derivative of $m(x)$ on which the average slope can be determined.

The CAPM is a linear model, thus it assumes that the power of the regression is one, so we take the power of the polynomial regression to be one and estimate the "beta" this way. Blundell (1991) shows that "beta" is simply the expected value of the derivative estimation; that is,

---

[5] We show it for the single variate regression; however, it can easily be generalized for multivariate case, (see Härdle et al., 2004).

$$\hat{\beta}^* = E\left(\hat{m}_h{}'(x)\right) \approx \frac{1}{n}\sum_{i=1}^{n}\hat{\beta}_1(X_i). \tag{23}$$

Eq. (23) is adequate for estimating market risk even if linearity does not hold. The advantage of this procedure is that it also considers cases if linearity is valid only over certain intervals; it is not necessary that the risk of a given asset is constant under any circumstances and it makes possible the estimation of extreme risk under extreme circumstances, thus this risk measure is more realistic than the simple CAPM beta.

Jensen (1968) alpha measure is biased if linearity does not hold, so we derive a semi-parametric measure similarly to the derivative estimation. The average performance of an asset can be determined by the surplus over its risk-adjusted return (calculated by the CAPM substituting the estimated semi-parametric beta) and this is called "alpha" or semi-parametric alpha; that is,

$$\hat{\alpha}^* = E\left(\hat{\alpha}^*(x)\right) \approx \frac{1}{n}\sum_{i=1}^{n}\left(Y_i - \hat{\beta}^* X_i\right), \tag{24}$$

where $\hat{\beta}^*$ is defined in Eq. (23). The Jensen alpha is an adequate performance measure only if the characteristic "curve" is over/under the theoretical CL (zero intercept) exactly by alpha at each point, which is true only if linearity holds. On the other hand, Eq. (24) estimates the abnormal return at each point which can vary from point to point, and the average performance is the mean of the point estimations.

## 3. Data

For the analysis we use 50-50 randomly selected stocks from the S&P 500, the S&P MidCap 400 and the S&P SmallCap 600 index components. These indexes represent the return of the large, the mid and the small capitalization stocks. The market return is the one available in the CRSP database which is capitalization weighted and adjusted with dividend. This index tracks the return of the New York Stock Exchange (NYSE), the American Stock Exchange (AMEX) and the NASDAQ stocks. Extending our analysis with the Fama-French three-factor model, we use the SMB and the HML factors from the CRSP database (the factors are based on six portfolios by size and book-to-market equity[6]). The risk-free rate is the return of the one-month Treasury bill from the CRSP. We use daily returns for a ten-year period from 1 January 1999 to 31 December 2008. Our data are not free of survivorship bias (see, e.g., Elton, 1996); that is, only those companies are eligible for insertion into the database which are still on the market at the end of the investigated period. This can introduce selection bias into the estimated parameters since those companies which go bankrupt, bear larger risk and might underperform the market significantly. However, we have to note that the survivorship bias is not a serious issue since we include mid and small cap firms besides large cap ones.

## 4. Results

In this section we present our results applying the estimations and test procedure presented in previous sections. First, we show the results of the single factor model, and then we present the generalization of the Fama-French three-factor (1996) model based on kernel regression. For presenting our results we choose one stock randomly, the Lowe's Companies Inc. (LOW), which is an S&P 500 component and another one which exhibits non-linearity in the CL, the National Oilwell Varco Inc (S&P 500 component).

---

[6] The detailed description of the portfolios can be found in Fama and French (1996) or on the website of Kenneth French.

*4.1. Single factor explanation of the US stock returns*

In this subsection we estimate single factor asset pricing models based on kernel regression. These estimates are the generalizations of the CAPM and our estimator allows alphas and betas to vary in time and the pricing function can be non-linear.

Figure 1 Panel A shows the CL of the Lowe's Inc. The bold curve is the kernel regression defined in Eq. (3), the dotted-dashed line is the linear regression defined in Eq. (2), and the dotted curves are the confidence band of the kernel regression at 95% level[7]. The $R^2$ of the kernel regression is almost 4% higher (0.369 vs. 0.356) and the alphas are significantly not different (they are the same up to four decimals); however, the linear beta is significantly downward biased (1.15 vs. 1.09). We cannot reject the linearity of the CL of the Lowe's Inc., the result of which is also supported by the 95% confidence band.[8] On the other hand, linearity of the National Oilwell Varco Inc. in Figure 1 Panel B can be rejected at 95% level.

**Please insert Figure 1 here**

We apply the above calculations for all the 150 randomly selected stocks in our database. The results are divided into three parts based on market capitalization: large,, mid and small cap. The average $R^2$ is higher for the kernel regression independently from firm size. Considering the large cap stocks, linearity of the CLs can be rejected in 9 cases out of 50 at 95% significance level. This result is significant on which we can reject the linear relation between the returns of S&P 500 components and the market return. In the case of mid and small capitalization stocks, we can reject linearity of the CL in two cases each which means the linearity of mid and small cap firms cannot be rejected at 95% confidence level. Altogether in 13 cases (8.7%) out of 150, we can reject the nullhypothesis; that is, linearity of the CLs of US stocks can be rejected at 95% confidence level (see Table 1).

---

[7] See Härdle (2004) for the calculation of the confidence band.
[8] We generate 250 different samples for the *T* test applying the wild bootstrapping method.

**Please insert Table 1 here**

We reject the linear nullhypothesis at 18% of the S&P 500 companies and at 8.7% of all the companies in our database, in which cases the parameters estimated by linear regression are biased and inconsistent, and thus the market risk ("beta") and the abnormal return ("alpha") should be estimated by non-linear methods. The parametric and semi-parametric alphas are not significantly different; however, the average difference between the linear and the kernel betas is 11% which is significant.[9] The linear CL of mid and small cap firms cannot be rejected, thus in these cases linear regression can be used. The abnormal returns of large cap stocks estimated by both methods do not differ significantly; the average alphas of those stocks which exhibit non-linearity are 0.05 vs. 0.05, and which exhibit linearity are 0.04 vs. 0.04. However, the average non-linear kernel betas are significantly higher than the OLS estimated (1.31 vs. 1.21), while if linearity holds there is not such a significant difference (0.90 vs. 0.92). There is no significant difference between the alphas of the mid cap stocks in the cases when linearity can be rejected (on average 0.14 vs. 0.13), and they are the same (0.05), in the cases when linearity cannot be rejected. Betas are different, if linearity does not hold, they are 1.72 vs. 1.39 on average; however, if we are not able to reject the linearity, there is no significant difference (0.93 vs. 0.94). In the case of small firms, the average alphas do not differ if linearity holds (0.07); however, betas are slightly different: 1.00 vs. 0.91. In the cases when linearity is rejetced, alphas are still the same on average (0.08) and betas do not differ significantly (1.33 vs. 1.34).

---

[9] The average difference is calculated by $\frac{1}{150}\sum_{j=1}^{150} abs\left(1 - \hat{\beta}_{LR} / \hat{\beta}_{KR}\right)$.

Summarizing the results of parameter estimations, we can argue that the kernel and the OLS alphas are almost the same; it does not matter whether or not linearity holds. On the other hand, the OLS beta is significantly downward biased when linearity of the CL can be rejected. The average beta is inversely related to firm size confirming the small firm effect (see, e.g., Banz, 1981; Basu, 1983). Those stocks which feature non-linear characteristic curve, have higher "betas" which is a sign that extremes cause invalidity of linearity since outliers raise the risk; on the other hand, the extremes are hard to be explained by the market movements linearly. It is a striking result that if linearity can be rejected, the difference in betas estimated by both methods is not the largest among small firms but among mid caps; however we have to note that there are only two mid cap stocks in the database whose characteristic curve exhibit non-linearity. If linearity can be rejected the difference is also significant at large cap stocks which along with the previous result indicate that linearity is not connected with firm size.

Figure 2 shows the derivative estimation of the characteristic curve of Lowe's Inc. in the function of the market risk premium. It can be seen that the estimated relevant risk is not constant; it exhibits high volatility at the tails, and in addition, on the positive tail, risk rises along with the market risk premium. Figure 2 Panel B shows the derivative estimation of the National Oilwell Varco Inc. The estimated market risk is similar to the one in Panel A; under normal circumstances, beta is constant; however, under extreme circumstances, estimation is very volatile. "Beta" is constant at the central part of the distribution maintaining the CAPM; however, at the tails it behaves differently. The non-constant risk estimation has several reasons: First, it can be imagined that linearity does not hold at the tails causing non-constant derivative estimation; second, the number of observations is low at the tails which introduces noise in the estimation.

**Please insert Figure 2 here**

Figure 3 Panel A shows the performance estimation of Lowe's Inc. in the function of the market risk premium. Under normal circumstances, "alpha" is stable and close to zero; however, security reacts differently to the extreme negative and positive movements of the market. The stock price can overreact the market movements on the extreme negative side inducing significant negative "alpha"; however, on the positive tail we measure significant positive performance; that is, the larger the extreme market movement, the larger the abnormal return is. We have to note that the regression estimate is less precise at the tails because of sparse data. Figure 3 Panel B shows the performance estimation of the National Oilwell Varco Inc. Results are similar to the case of the Lowe's Inc., under normal circumstances, the "alpha" is relatively stable and significantly not different from zero; while under extreme circumstances, the performance is significantly negative at the negative tail and significantly positive on the positive tail. The results of the semi-parametric alpha estimations induce an important notice for the mutual fund industry: managers can beat the market only when extreme market movements occur.

**Please insert Figure 3 here**

We use the semi-parametric betas in Table 1 as explanatory variable and the average daily returns as the dependent variable to estimate the cross sectional regressions (SMLs). Table 2 shows the estimated parameters of the SMLs by market capitalization.

**Please insert Table 2 here**

The mid cap SML has the highest slope coefficient (0.0407) as the slope of the large cap line is very small (0.0081) and the curve is almost flat. The slope of the SML estimated for S&P SmallCap 600 stocks is negative (-0.0344); that is, larger risk would associate with smaller expected return. This result shows that the the single factor model cannot explain the small firm effect (see, e.g., Banz, 1981; Basu, 1983; Fama and French, 1995; 1996). The SML estimated for all the 150 companies in our database has a slope coefficient of 0.0085 which is still very flat. Figure 4 shows the SMLs for the S&P size indexes and for all the stocks. We apply the wild bootstrap linearity test for the SMLs and we can conclude that linearity cannot be rejected at any usual significance level. Based on the cross sectional regressions, we can argue that neither time-varying betas, nor the non-linear pricing function can explain the cross sections of US stock returns in a single factor setting as kernel regressions are not able to explain the small firm effect.

**Please insert Figure 4 here**

*4.2. Multifactor explanation of the US stock returns*

As we have shown in the previous subsection, the single variate asset pricing model cannot explain the US stock returns because it fails to describe the returns of small firms, thus we estimate the linear and the kernel versions of the Fama-French three-factor model. The Fama-French three-factor-model (Fama and French, 1996) extends the single factor model adding two risk factors: one for the small firm effect which is the return spread between the small and the large capitalization stocks (SMB) and one for the 'relative distress' which is the return spread between the high and the low book-to-market equity fimrs (HML).

Table 3 shows the result of the linear and the kernel regressions based on the Fama-French factors. The kernel form of the Fama-French three-factor model is estimated by Eq. (4) and the semi-parametric $\hat{\beta}_{KR}$, $\hat{s}_{KR}$ and $\hat{h}_{KR}$ are the simple average of the factors estimated at each data point in a similar manner presented in Eq. (23), and the semi-parametric alpha ($\hat{\alpha}_{KR}$) is estimated by the following equation:

$$\hat{\alpha}_{KR} = E\left(\hat{\alpha}_{KR}(x)\right) \approx \frac{1}{n}\sum_{i=1}^{n}\left(Y_i - \hat{\beta}_{KR}X_{1,i} - \hat{s}_{KR}X_{2,i} - \hat{h}_{KR}X_{3,i}\right) \qquad (25)$$

where $X_d$s are the regressors (the market risk premium, the SMB and the HML factor, respectively).

However, linearity can be rejected in none of the cases as the $R^2$s are higher for the kernel regressions in all but one case. Our semi-parametric measures are only slightly different from the parameter estimations which is not striking since the linearity cannot be rejected. The betas are positively related to the expected return even among small cap stocks, the correlation coefficient between the small cap kernel betas and the small cap average returns is 10.55%. The small cap stocks load most heavily on the SMB factor, then the mid cap stocks and the least upon the large cap stocks. The three factors are enough to explain the expected returns of US stocks even if linear regression is applied; however, a single factor model cannot explain the cross sections of US stock returns not even if we allow time-varying betas and non-linear pricing functions.

**Please insert Table 3 here**

Thus, if the three-factor model is applied, the non-parametric approach adds no value with respect to the linear model. However, the betas of both the single factor and the three-factor models have the same positive sign even in the cases of small cap stocks. Additionally, the linear and the kernel CAPM betas exhibit negative relationship with the expected returns, while either the linear or the kernel betas in a three factor setting correlate positively with the expected returns.

## 5. Concluding remarks

Based on our results, linearity of the CL applying the single factor CAPM can be rejected in 8.7% of the US stocks, so the standard linear estimators cannot be used for estimating market risk and performance. We propose using semi-parametric approaches for estimating "alphas" and "betas" since our study shows that they do not differ significantly from the linear measures when linearity holds, while they provide good alternatives when linearity can be rejected. Both risk and performance measures; that is, beta and alpha, are not constant when extreme market movements occur; however, standard measures would indicate constant estimations. The linear beta in the single factor setting is significantly downward biased when linearity does not hold, which is a problem that can be solved by our semi-parametric beta measure. Our semi-parametric alpha shows that managers can beat the market only under extreme market conditions. On the other hand, the single factor CAPM cannot be defended because it fails to explain – with either linear or non-linear pricing function or with either constant or non-constant beta – the retun of small firms.

Altogether, the one-factor model fails to explain the cross section of small cap stock returns; however, the generalized Fama-French three-factor model based on kernel regression holds; furthermore, it holds even in a linear setting and it is also able to explain the returns of small cap stocks.

**Acknowledgements**

We thank Laci Györfi for very useful comments and suggestions and we are grateful to the seminar and conference participants at Masaryk University, University of Zagreb and the Morgan Stanley, Budapest. We would like to gratefully acknowledge the valuable comments of the anonymus referee which contribute to a better understanding of this paper. This work is connected to the scientific program of the "Development of quality-oriented and harmonized R+D+I strategy and functional model at BME" project. This project is supported by the New Hungary Development Plan (Project ID: TÁMOP-4.2.1/B-09/1/KMR-2010-0002).

**Figure titles**

Figure 1. The characteristic curve of the Lowe's Companies Inc. (Panel A) and the National Oilwell Varco Inc. (Panel B)

Figure 2. Semi-parametric derivative estimation

Figure 3. Semi-parametric alpha estimation

Figure 4. Security market lines for the S&P 500, the S&P MidCap 400, the S&P SmallCap 600 and for all the stocks

Table 1 The results of kernel and linear regressions of characteristic curves

| | | | | Panel A - S&P 500 | | | | | | |
|---|---|---|---|---|---|---|---|---|---|---|
| Ticker | No. of obs. | E(r) | p-value | h | $R^2_{KR}$ | $\hat{\alpha}_{KR}$ | $\hat{\beta}_{KR}$ | $R^2_{LR}$ | $\hat{\alpha}_{LR}$ | $\hat{\beta}_{LR}$ |
| ACAS | 2515 | 0.01 | 0.26 | 0.39 | 0.34 | 0.00 | 0.92 | 0.29 | 0.00 | 1.09 |
| AES | 2515 | 0.05 | 0.52 | 0.58 | 0.15 | 0.04 | 1.13 | 0.14 | 0.04 | 1.17 |
| APH | 2515 | 0.11 | 0.50 | 0.42 | 0.34 | 0.10 | 1.19 | 0.33 | 0.10 | 1.18 |
| BA | 2515 | 0.04 | 0.25 | 0.43 | 0.29 | 0.03 | 0.82 | 0.27 | 0.03 | 0.83 |
| BAX | 2515 | 0.05 | *0.01 | 0.49 | 0.15 | 0.04 | 0.55 | 0.13 | 0.04 | 0.52 |
| BJS | 2515 | 0.09 | 0.28 | 0.84 | 0.19 | 0.09 | 0.85 | 0.17 | 0.09 | 0.98 |
| BMC | 2515 | 0.04 | **0.02 | 0.40 | 0.25 | 0.04 | 1.49 | 0.24 | 0.04 | 1.30 |
| BRL | 2509 | 0.10 | 0.31 | 0.44 | 0.10 | 0.09 | 0.65 | 0.09 | 0.09 | 0.63 |
| CPWR | 2515 | 0.01 | **0.05 | 0.40 | 0.23 | 0.01 | 1.50 | 0.22 | 0.01 | 1.41 |
| D | 2515 | 0.05 | 0.56 | 0.70 | 0.19 | 0.04 | 0.43 | 0.19 | 0.04 | 0.49 |
| DD | 2515 | 0.00 | 0.36 | 0.41 | 0.36 | -0.01 | 0.84 | 0.35 | -0.01 | 0.86 |
| DHI | 2515 | 0.07 | 0.42 | 0.36 | 0.30 | 0.06 | 1.53 | 0.29 | 0.06 | 1.34 |
| DOV | 2515 | 0.02 | 0.36 | 0.38 | 0.44 | 0.01 | 1.01 | 0.44 | 0.01 | 0.97 |
| FDX | 2515 | 0.04 | **0.04 | 0.39 | 0.31 | 0.03 | 0.93 | 0.30 | 0.03 | 0.87 |
| FO | 2515 | 0.04 | 0.14 | 0.39 | 0.28 | 0.03 | 0.66 | 0.26 | 0.03 | 0.66 |
| GM | 2515 | -0.05 | 0.47 | 0.77 | 0.29 | -0.06 | 1.13 | 0.28 | -0.06 | 1.29 |
| HCBK | 2383 | 0.11 | 0.64 | 0.79 | 0.26 | 0.10 | 0.55 | 0.26 | 0.10 | 0.61 |
| HCP | 2515 | 0.08 | *0.06 | 0.66 | 0.32 | 0.07 | 0.64 | 0.27 | 0.07 | 0.82 |
| HPC | 2483 | 0.02 | 0.27 | 0.37 | 0.23 | 0.02 | 0.95 | 0.20 | 0.02 | 0.93 |
| HSP | 1175 | 0.02 | 0.16 | 0.37 | 0.24 | 0.01 | 0.67 | 0.21 | 0.01 | 0.62 |
| KFT | 1898 | 0.01 | 0.26 | 0.49 | 0.22 | 0.00 | 0.39 | 0.20 | 0.00 | 0.49 |
| LMT | 2515 | 0.05 | 0.48 | 0.61 | 0.13 | 0.04 | 0.44 | 0.12 | 0.04 | 0.51 |
| LOW | 2515 | 0.05 | 0.20 | 0.41 | 0.37 | 0.04 | 1.15 | 0.36 | 0.04 | 1.09 |
| LSI | 2515 | 0.05 | *0.06 | 0.39 | 0.33 | 0.05 | 1.91 | 0.31 | 0.05 | 1.76 |
| LXK | 2515 | 0.02 | ***0.00 | 0.41 | 0.18 | 0.02 | 1.09 | 0.17 | 0.02 | 0.94 |
| MDT | 2515 | 0.01 | 0.72 | 0.43 | 0.21 | 0.00 | 0.61 | 0.20 | 0.00 | 0.64 |
| MI | 2515 | 0.02 | 0.19 | 0.55 | 0.40 | 0.01 | 0.93 | 0.35 | 0.01 | 1.10 |
| MTW | 2515 | 0.05 | 0.60 | 0.60 | 0.32 | 0.05 | 1.21 | 0.31 | 0.05 | 1.25 |
| MUR | 2515 | 0.09 | *0.07 | 0.71 | 0.23 | 0.08 | 0.60 | 0.20 | 0.08 | 0.77 |
| MWW | 2515 | 0.08 | 0.12 | 0.39 | 0.30 | 0.07 | 1.98 | 0.29 | 0.07 | 1.77 |
| NOV | 2515 | 0.12 | **0.04 | 0.38 | 0.26 | 0.11 | 1.04 | 0.22 | 0.11 | 1.21 |
| NUE | 2515 | 0.11 | 0.30 | 0.39 | 0.36 | 0.10 | 1.25 | 0.34 | 0.10 | 1.27 |
| ORCL | 2515 | 0.09 | **0.02 | 0.40 | 0.36 | 0.09 | 1.59 | 0.35 | 0.09 | 1.49 |
| PAYX | 2515 | 0.04 | ***0.00 | 0.39 | 0.32 | 0.03 | 1.11 | 0.30 | 0.03 | 1.01 |
| PG | 2515 | 0.03 | 0.21 | 0.56 | 0.15 | 0.02 | 0.43 | 0.13 | 0.02 | 0.44 |
| QCOM | 2515 | 0.16 | 0.24 | 0.39 | 0.31 | 0.16 | 1.72 | 0.30 | 0.16 | 1.51 |
| S | 2515 | -0.06 | *0.06 | 0.53 | 0.29 | -0.07 | 1.01 | 0.25 | -0.06 | 1.25 |
| S | 1565 | 0.06 | 0.60 | 0.28 | 0.17 | 0.04 | 0.90 | 0.16 | 0.04 | 0.90 |
| SLE | 2515 | -0.01 | 0.14 | 0.55 | 0.21 | -0.02 | 0.50 | 0.18 | -0.02 | 0.56 |
| SPG | 2515 | 0.07 | 0.28 | 0.70 | 0.37 | 0.06 | 0.68 | 0.30 | 0.06 | 0.90 |
| SRE | 2515 | 0.05 | 0.31 | 0.51 | 0.24 | 0.04 | 0.55 | 0.21 | 0.04 | 0.62 |
| SYMC | 2515 | 0.12 | **0.02 | 0.40 | 0.22 | 0.12 | 1.27 | 0.21 | 0.12 | 1.19 |
| TDC | 316 | -0.15 | 0.32 | 0.63 | 0.46 | -0.04 | 0.81 | 0.42 | -0.04 | 0.85 |
| TE | 2515 | 0.01 | 0.25 | 0.54 | 0.19 | 0.00 | 0.58 | 0.16 | 0.00 | 0.61 |
| TGT | 2515 | 0.04 | 0.39 | 0.39 | 0.36 | 0.03 | 1.09 | 0.34 | 0.03 | 1.06 |
| TLAB | 2515 | -0.01 | ***0.00 | 0.39 | 0.30 | -0.02 | 1.76 | 0.28 | -0.02 | 1.50 |
| WFR | 2515 | 0.18 | 0.45 | 0.36 | 0.17 | 0.17 | 1.77 | 0.16 | 0.17 | 1.75 |
| WLP | 1805 | 0.06 | 0.21 | 0.36 | 0.24 | 0.05 | 0.66 | 0.19 | 0.05 | 0.68 |
| WLP | 1486 | 0.10 | 0.72 | 0.65 | 0.06 | 0.08 | 0.45 | 0.05 | 0.08 | 0.42 |
| ZMH | 1860 | 0.03 | 0.22 | 0.38 | 0.22 | 0.03 | 0.69 | 0.20 | 0.03 | 0.65 |
| Average | 2362 | 0.05 | - | 0.48 | 0.26 | 0.04 | 0.97 | 0.24 | 0.04 | 0.98 |

*Panel B - S&P MidCap 400*

| Ticker | No. of obs. | E(r) | p-value | h | $R^2_{KR}$ | $\hat{\alpha}^J_{KR}$ | $\hat{\beta}^J_{KR}$ | $R^2_{LR}$ | $\hat{\alpha}^J_{LR}$ | $\hat{\beta}^J_{LR}$ |
|---|---|---|---|---|---|---|---|---|---|---|
| AAI | 2515 | 0.11 | 0.60 | 0.63 | 0.17 | 0.10 | 1.41 | 0.16 | 0.10 | 1.24 |
| AMG | 2484 | 0.07 | 0.34 | 0.57 | 0.45 | 0.06 | 1.32 | 0.43 | 0.06 | 1.38 |
| ARG | 2515 | 0.10 | 0.49 | 0.48 | 0.24 | 0.09 | 0.99 | 0.23 | 0.09 | 1.03 |
| AVCT | 2118 | 0.03 | *0.06 | 0.53 | 0.26 | 0.04 | 1.52 | 0.25 | 0.04 | 1.44 |
| BRO | 2515 | 0.08 | 0.21 | 0.47 | 0.21 | 0.07 | 0.65 | 0.20 | 0.07 | 0.64 |
| CLF | 2515 | 0.13 | *0.06 | 0.40 | 0.32 | 0.13 | 1.13 | 0.26 | 0.13 | 1.35 |
| CMG | 737 | 0.10 | ***0.00 | 0.38 | 0.29 | 0.14 | 1.40 | 0.24 | 0.13 | 1.00 |
| CPT | 2515 | 0.05 | 0.27 | 0.83 | 0.32 | 0.04 | 0.60 | 0.27 | 0.04 | 0.76 |
| CR | 2515 | 0.01 | 0.29 | 0.31 | 0.39 | 0.00 | 0.94 | 0.33 | 0.00 | 0.92 |
| CWTR | 2515 | 0.12 | 0.16 | 0.36 | 0.15 | 0.12 | 1.31 | 0.13 | 0.12 | 1.25 |
| CXW | 2515 | 0.05 | 0.51 | 0.62 | 0.07 | 0.04 | 0.71 | 0.06 | 0.04 | 0.69 |
| DCI | 2515 | 0.07 | 0.68 | 0.43 | 0.31 | 0.06 | 0.82 | 0.31 | 0.06 | 0.84 |
| ELY | 2515 | 0.04 | 0.41 | 0.46 | 0.22 | 0.04 | 1.02 | 0.21 | 0.04 | 0.98 |
| ENR | 2168 | 0.07 | 0.12 | 0.54 | 0.18 | 0.07 | 0.71 | 0.17 | 0.07 | 0.71 |
| FMER | 2515 | 0.03 | 0.38 | 0.33 | 0.40 | 0.02 | 0.96 | 0.35 | 0.02 | 1.01 |
| FNFG | 2363 | 0.09 | 0.30 | 0.53 | 0.24 | 0.08 | 0.74 | 0.22 | 0.08 | 0.77 |
| FTO | 2515 | 0.15 | *0.10 | 0.37 | 0.22 | 0.15 | 0.92 | 0.18 | 0.15 | 1.09 |
| HBI | 581 | -0.03 | 0.32 | 0.69 | 0.32 | 0.01 | 0.91 | 0.28 | 0.01 | 1.00 |
| HE | 2515 | 0.03 | 0.89 | 0.62 | 0.18 | 0.02 | 0.41 | 0.18 | 0.02 | 0.39 |
| HMA | 2515 | -0.02 | *0.08 | 0.82 | 0.11 | -0.03 | 0.51 | 0.06 | -0.03 | 0.61 |
| HNI | 2515 | 0.02 | 0.52 | 0.74 | 0.25 | 0.01 | 0.84 | 0.25 | 0.01 | 0.93 |
| HRC | 2515 | 0.00 | 0.35 | 0.45 | 0.14 | -0.01 | 0.47 | 0.13 | -0.01 | 0.50 |
| HRL | 2515 | 0.05 | 0.25 | 0.36 | 0.11 | 0.04 | 0.46 | 0.09 | 0.04 | 0.38 |
| IDXX | 2515 | 0.07 | 0.63 | 0.46 | 0.16 | 0.06 | 0.77 | 0.16 | 0.06 | 0.75 |
| IEX | 2515 | 0.06 | 0.68 | 0.40 | 0.35 | 0.05 | 0.90 | 0.34 | 0.05 | 0.90 |
| IRF | 2515 | 0.08 | *0.07 | 0.35 | 0.31 | 0.08 | 1.71 | 0.29 | 0.08 | 1.51 |
| JBHT | 2515 | 0.10 | 0.32 | 0.43 | 0.23 | 0.10 | 1.06 | 0.22 | 0.10 | 1.01 |
| JBLU | 1673 | 0.02 | 0.67 | 0.52 | 0.21 | 0.00 | 1.22 | 0.21 | 0.00 | 1.14 |
| KMT | 2515 | 0.06 | 0.10 | 0.41 | 0.38 | 0.06 | 1.04 | 0.37 | 0.06 | 1.09 |
| LRCX | 2515 | 0.14 | **0.02 | 0.38 | 0.35 | 0.13 | 2.04 | 0.32 | 0.13 | 1.77 |
| MEG | 2515 | -0.06 | 0.26 | 1.05 | 0.16 | -0.07 | 0.82 | 0.15 | -0.07 | 1.07 |
| MRX | 2515 | 0.02 | 0.31 | 0.42 | 0.17 | 0.01 | 0.77 | 0.16 | 0.01 | 0.80 |
| MTX | 2515 | 0.02 | 0.58 | 0.44 | 0.32 | 0.01 | 0.82 | 0.32 | 0.01 | 0.84 |
| MVL | 2515 | 0.14 | 0.41 | 0.40 | 0.09 | 0.13 | 0.76 | 0.09 | 0.13 | 0.79 |
| NHP | 2515 | 0.07 | 0.14 | 0.54 | 0.31 | 0.06 | 0.77 | 0.29 | 0.06 | 0.83 |
| OII | 2515 | 0.10 | 0.15 | 0.53 | 0.18 | 0.09 | 0.86 | 0.16 | 0.10 | 0.94 |
| ORLY | 2515 | 0.07 | *0.08 | 0.38 | 0.22 | 0.06 | 0.94 | 0.20 | 0.06 | 0.89 |
| OSK | 2515 | 0.07 | 0.65 | 0.49 | 0.20 | 0.06 | 0.94 | 0.18 | 0.06 | 0.94 |
| PSD | 2515 | 0.03 | 0.43 | 0.63 | 0.13 | 0.02 | 0.41 | 0.13 | 0.02 | 0.38 |
| RS | 2515 | 0.08 | *0.08 | 0.44 | 0.37 | 0.07 | 1.12 | 0.34 | 0.07 | 1.29 |
| RYL | 2515 | 0.09 | 0.53 | 0.45 | 0.29 | 0.08 | 1.40 | 0.28 | 0.08 | 1.27 |
| SKS | 2515 | -0.01 | 0.32 | 0.39 | 0.22 | -0.01 | 1.09 | 0.19 | -0.01 | 1.05 |
| SRCL | 2515 | 0.14 | 0.44 | 0.39 | 0.11 | 0.13 | 0.67 | 0.10 | 0.13 | 0.64 |
| TECD | 2515 | 0.01 | 0.27 | 0.40 | 0.22 | 0.01 | 1.19 | 0.20 | 0.01 | 1.02 |
| THG | 2515 | 0.02 | 0.40 | 0.37 | 0.23 | 0.02 | 0.93 | 0.22 | 0.02 | 0.93 |
| UTHR | 2347 | 0.15 | 0.26 | 0.34 | 0.09 | 0.16 | 0.97 | 0.08 | 0.16 | 0.83 |
| UTR | 2515 | 0.01 | 0.28 | 0.50 | 0.36 | 0.00 | 0.82 | 0.32 | 0.00 | 0.92 |
| VARI | 2406 | 0.10 | 0.32 | 0.33 | 0.25 | 0.11 | 1.27 | 0.23 | 0.11 | 1.16 |
| WBS | 2515 | 0.01 | 0.14 | 0.44 | 0.39 | 0.00 | 0.93 | 0.37 | 0.00 | 1.00 |
| WOR | 2515 | 0.05 | 0.18 | 0.41 | 0.32 | 0.04 | 1.12 | 0.30 | 0.04 | 1.15 |
| Average | 2400 | 0.06 | - | 0.48 | 0.24 | 0.06 | 0.96 | 0.22 | 0.06 | 0.96 |

|  |  |  |  |  | Panel C - S&P SmallCap 600 |  |  |  |  |  |
|---|---|---|---|---|---|---|---|---|---|---|
| Ticker | No. of obs. | E(r) | p-value | h | $R^2_{KR}$ | $\hat{\alpha}_{KR}$ | $\hat{\beta}_{KR}$ | $R^2_{LR}$ | $\hat{\alpha}_{LR}$ | $\hat{\beta}_{LR}$ |
| ABM | 2454 | 0.06 | 0.38 | 0.40 | 0.25 | 0.03 | 0.85 | 0.23 | 0.03 | 0.80 |
| ACLS | 2077 | 0.02 | 0.63 | 0.54 | 0.27 | -0.03 | 2.20 | 0.26 | -0.04 | 1.84 |
| CASY | 2426 | 0.04 | *0.06 | 0.44 | 0.23 | 0.06 | 1.05 | 0.21 | 0.06 | 0.86 |
| CCRN | 1766 | 0.04 | 0.28 | 0.41 | 0.19 | 0.00 | 1.25 | 0.18 | 0.00 | 0.97 |
| DNEX | 2473 | 0.06 | 0.39 | 0.50 | 0.20 | 0.04 | 0.95 | 0.20 | 0.04 | 0.86 |
| EE | 2373 | 0.03 | 0.42 | 0.54 | 0.21 | 0.05 | 0.74 | 0.21 | 0.05 | 0.66 |
| GFF | 2396 | -0.01 | 0.38 | 0.51 | 0.20 | 0.03 | 0.92 | 0.20 | 0.03 | 0.90 |
| GTIV | 2142 | -0.01 | 0.31 | 0.40 | 0.15 | 0.16 | 0.82 | 0.13 | 0.16 | 0.70 |
| HMSY | 2313 | 0.05 | 0.37 | 0.90 | 0.03 | 0.13 | 0.46 | 0.03 | 0.13 | 0.51 |
| HOMB | 627 | 0.09 | 0.36 | 0.75 | 0.30 | 0.13 | 1.06 | 0.28 | 0.13 | 0.81 |
| HTLD | 2440 | 0.09 | 0.11 | 0.40 | 0.21 | 0.08 | 1.02 | 0.20 | 0.09 | 0.85 |
| INT | 2412 | 0.04 | 0.30 | 0.77 | 0.23 | 0.13 | 0.79 | 0.19 | 0.12 | 1.02 |
| ISYS | 2405 | 0.06 | 0.96 | 0.56 | 0.11 | 0.07 | 0.87 | 0.11 | 0.07 | 0.86 |
| KNOT | 1292 | 0.05 | 0.84 | 0.91 | 0.04 | -0.03 | 0.68 | 0.04 | -0.03 | 0.85 |
| MANT | 1716 | 0.02 | 0.24 | 0.48 | 0.11 | 0.11 | 0.94 | 0.10 | 0.11 | 0.71 |
| MFB | 850 | 0.13 | 0.65 | 0.57 | 0.31 | -0.01 | 1.25 | 0.30 | -0.01 | 1.19 |
| MNT | 2461 | 0.09 | 0.28 | 3.31 | 0.01 | 0.09 | 0.68 | 0.04 | 0.09 | 0.48 |
| MNT | 398 | -0.01 | 0.30 | 1.01 | 0.01 | 0.02 | 0.23 | 0.01 | 0.02 | 0.19 |
| MOH | 1366 | 0.05 | 0.22 | 0.90 | 0.10 | 0.03 | 0.61 | 0.10 | 0.03 | 0.70 |
| MTH | 2417 | 0.10 | 0.27 | 0.70 | 0.25 | 0.14 | 1.56 | 0.21 | 0.14 | 1.48 |
| NILE | 1152 | 0.14 | 0.25 | 0.57 | 0.16 | 0.05 | 1.32 | 0.14 | 0.05 | 0.99 |
| NOVN | 2440 | 0.12 | 0.52 | 0.50 | 0.09 | 0.12 | 1.04 | 0.08 | 0.12 | 0.94 |
| NPK | 2431 | 0.07 | 0.24 | 0.67 | 0.28 | 0.06 | 0.57 | 0.26 | 0.06 | 0.65 |
| NSIT | 2478 | 0.05 | 0.65 | 0.43 | 0.19 | 0.03 | 1.44 | 0.18 | 0.03 | 1.31 |
| NTRI | 1327 | 0.13 | 0.25 | 0.56 | 0.05 | 0.05 | 1.03 | 0.04 | 0.05 | 0.78 |
| NWK | 2404 | 0.10 | 0.35 | 0.54 | 0.11 | 0.04 | 1.32 | 0.11 | 0.04 | 1.09 |
| ONB | 2387 | 0.10 | 0.64 | 0.67 | 0.26 | 0.02 | 0.77 | 0.25 | 0.02 | 0.75 |
| PBY | 2428 | 0.00 | 0.31 | 0.35 | 0.21 | 0.01 | 1.30 | 0.19 | 0.01 | 1.16 |
| PEI | 2397 | 0.04 | **0.03 | 0.86 | 0.34 | 0.01 | 0.75 | 0.28 | 0.01 | 1.01 |
| PENX | 2396 | 0.06 | 0.35 | 0.64 | 0.08 | 0.04 | 0.65 | 0.08 | 0.04 | 0.72 |
| PRFT | 2228 | 0.05 | 0.62 | 0.65 | 0.04 | 0.22 | 0.94 | 0.03 | 0.22 | 1.01 |
| PRGS | 2461 | 0.09 | 0.52 | 0.40 | 0.20 | 0.04 | 1.08 | 0.20 | 0.04 | 0.99 |
| PVTB | 2255 | 0.14 | 0.67 | 0.54 | 0.12 | 0.10 | 0.74 | 0.11 | 0.10 | 0.65 |
| RADS | 2439 | 0.10 | 0.49 | 0.40 | 0.17 | 0.09 | 1.39 | 0.15 | 0.09 | 1.41 |
| RBN | 2424 | 0.22 | *0.08 | 0.41 | 0.21 | 0.05 | 0.90 | 0.19 | 0.05 | 0.89 |
| SAFM | 2402 | 0.16 | *0.08 | 0.35 | 0.11 | 0.10 | 0.79 | 0.09 | 0.10 | 0.67 |
| SKT | 2426 | -0.06 | 0.14 | 0.56 | 0.29 | 0.09 | 0.68 | 0.26 | 0.09 | 0.72 |
| SLXP | 1997 | 0.10 | 0.28 | 0.58 | 0.09 | 0.09 | 1.15 | 0.08 | 0.09 | 0.90 |
| SNS | 2397 | 0.00 | 0.61 | 0.64 | 0.15 | -0.01 | 0.96 | 0.15 | -0.01 | 0.82 |
| SNS | 106 | 0.11 | 0.44 | 1.21 | 0.05 | 0.27 | 0.51 | 0.06 | 0.27 | 0.57 |
| SSS | 2434 | 0.04 | 0.36 | 0.55 | 0.34 | 0.06 | 0.68 | 0.31 | 0.06 | 0.78 |
| SUP | 2464 | 0.04 | 0.54 | 0.49 | 0.26 | -0.01 | 0.87 | 0.25 | -0.01 | 0.82 |
| SYMM | 2432 | -0.05 | 0.32 | 0.47 | 0.17 | 0.08 | 1.53 | 0.16 | 0.08 | 1.33 |
| TSFG | 2446 | 0.05 | 0.13 | 0.38 | 0.31 | -0.02 | 1.06 | 0.24 | -0.02 | 1.20 |
| TXRH | 1046 | 0.00 | 0.14 | 0.32 | 0.30 | 0.00 | 1.43 | 0.20 | 0.00 | 0.87 |
| UEIC | 2426 | -0.02 | 0.48 | 0.40 | 0.16 | 0.09 | 1.13 | 0.16 | 0.09 | 1.02 |
| UNS | 2381 | 0.10 | 0.19 | 0.52 | 0.21 | 0.05 | 0.54 | 0.20 | 0.05 | 0.61 |
| VECO | 2482 | 0.02 | 0.21 | 0.46 | 0.28 | -0.01 | 1.88 | 0.24 | -0.01 | 1.56 |
| VSEA | 2429 | 0.20 | **0.03 | 0.32 | 0.30 | 0.14 | 1.90 | 0.27 | 0.14 | 1.67 |
| ZEP | 287 | 0.30 | 0.54 | 1.22 | 0.39 | 0.40 | 1.21 | 0.38 | 0.38 | 1.07 |
| Average | 2052 | 0.07 | - | 0.63 | 0.19 | 0.07 | 1.01 | 0.17 | 0.07 | 0.92 |

*Notes:* In Panel A,B,C the estimated parameters and statistics of characteristic curves of 50-50-50 randomly chosen companies from the S&P 500, S&P MidCap 400 and the S&P SmallCap 600 universe, can be seen respectively. We estimate the characteristic curve of all the companies by two different methods: 1, by the kernel regression, the non-parametric CAPM is estimated, that is $r_j - r_f = \hat{m}_h(r_m - r_f) + \hat{\eta}_j$ where $\hat{\eta}_j$ are the residuals, $r_j$-$r_f$ and $r_M$-$r_f$ are the risk premium of the stock and the market respectively, $h$ is an optimally selected bandwidth by Cross Validation (CV) (We use the Gaussian kernel and the Nadaraya (1964) and Watson (1964) weighting function in the kernel

regression); 2, by linear regression, that is $r_j - r_f = \hat{\alpha}_j + \hat{\beta}_j (r_m - r_f) + \hat{\varepsilon}_j$ is estimated where $\hat{\varepsilon}_j$ is the residual series. The market return is the Center for Research in Security Prices (CRSP) value weighted index return; the risk-free rate is the return of the one-month Treasury bill. In the first column can be seen the tickers of each stock in the database, in column 2 the number of observations, in column 3 the mean return for the estimation period, in column 4 the *p*-value of the linearity test, in column 5 the optimal bandwidth of the kernel regression and then the $R^2$s, alphas and betas of the kernel and the linear regressions. The number of asterisks means rejection of the linear nullhypothesis at 90%, 95%, 99% confidence level.

Table 2 Estimations of SMLs

| | | | *Security Market Lines* | | | | | | | |
|---|---|---|---|---|---|---|---|---|---|---|
| Segment | No. of obs. | E(r) | p-value | h | $R^2_{KR}$ | $\hat{\alpha}'_{KR}$ | $\hat{\beta}'_{KR}$ | $R^2_{LR}$ | $\hat{\alpha}'_{LR}$ | $\hat{\beta}'_{LR}$ |
| S&P 500 | 50 | 0.04798 | 0.144 | 20.504 | 0.1169 | 0.0401 | 0.0081 | 0.0516 | 0.0192 | 0.0296 |
| S&P MidCap 400 | 50 | 0.06083 | 0.728 | 15.896 | 0.1311 | 0.0217 | 0.0407 | 0.1043 | 0.0141 | 0.0486 |
| S&P SmallCap 600 | 50 | 0.06749 | 0.760 | 18.728 | 0.0885 | 0.1022 | -0.0344 | 0.0263 | 0.0957 | -0.0280 |
| All companies | 150 | 0.05877 | 0.688 | 14.751 | 0.0424 | 0.0504 | 0.0085 | 0.0100 | 0.0439 | 0.0152 |

*Notes:* We estimate SMLs for the S&P 500, S&P MidCap 400, S&P SmallCap 600 stocks and for all the stocks in our database. The rows 1,2,3,4 include the estimated parameters and statistics of the SMLs for the S&P 500, the S&P MidCap 400, the S&P SmallCap 600 and for all the stocks, respectively. We estimate the SMLs by two methods: 1, by the kernel regressions in the form $\overline{r}_j = \hat{m}_h(\hat{\beta}^*_j) + \hat{\lambda}_j$, where $\overline{r}_j$ is the average return of stock *j* in the investigated period, $\hat{\beta}^*_j$ is the semi-parametric beta of stock *j*, $\hat{\lambda}_j$ is the residual series and *h* is an optimally selected bandwidth by Cross Validation (CV) (We use the Gaussian kernel and the Nadaraya (1964) and Watson (1964) weighting function in the kernel regression.); 2, by linear regression in the form $\overline{r}_j = \hat{\alpha}'_j + \hat{\beta}'_j \hat{\beta}^*_j + \hat{\kappa}_j$, where $\hat{\kappa}_j$ are the residuals. The market return is the Center for Research in Security Prices (CRSP) value weighted index return; the risk-free rate is the return of the one-month Treasury bill. In the third column can be seen the mean return of the given index for the estimation period, in column 4 the *p*-value of the linearity test, in column 5 the optimal bandwidth of the kernel regression and then the $R^2$s, alphas and betas of the kernel and the linear regressions.

Table 3 The results of kernel and linear regressions of characteristic curves

| | | | | | | Panel A - S&P 500 | | | | | |
|---|---|---|---|---|---|---|---|---|---|---|---|
| Ticker | p-value | $R^2_{KR}$ | $\hat{\alpha}_{KR}$ | $\hat{\beta}_{KR}$ | $\hat{s}_{KR}$ | $\hat{h}_{KR}$ | $R^2_{LR}$ | $\hat{\alpha}_{LR}$ | $\hat{\beta}_{LR}$ | $\hat{s}_{LR}$ | $\hat{h}_{LR}$ |
| ACAS | 0.40 | 0.464 | -0.014 | 0.888 | 0.003 | 0.004 | 0.348 | -0.027 | 1.244 | 0.003 | 0.010 |
| AES | 0.35 | 0.249 | 0.013 | 1.313 | 0.004 | 0.009 | 0.146 | 0.027 | 1.266 | 0.000 | 0.006 |
| APH | 0.32 | 0.442 | 0.091 | 1.114 | 0.008 | -0.001 | 0.358 | 0.098 | 1.161 | 0.006 | -0.002 |
| BA | 0.22 | 0.396 | 0.025 | 0.974 | 0.000 | 0.003 | 0.288 | 0.026 | 0.876 | -0.002 | 0.003 |
| BAX | 0.13 | 0.305 | 0.044 | 0.608 | -0.002 | -0.001 | 0.154 | 0.039 | 0.543 | -0.004 | 0.002 |
| BJS | 0.45 | 0.270 | 0.046 | 1.190 | 0.004 | 0.014 | 0.200 | 0.064 | 1.111 | 0.002 | 0.008 |
| BMC | 0.51 | 0.330 | 0.053 | 1.190 | 0.004 | -0.010 | 0.275 | 0.052 | 1.168 | 0.003 | -0.009 |
| BRL | 0.32 | 0.192 | 0.084 | 0.532 | 0.004 | -0.002 | 0.099 | 0.087 | 0.618 | 0.002 | -0.001 |
| CPWR | 0.29 | 0.384 | 0.010 | 1.330 | 0.007 | -0.007 | 0.241 | 0.015 | 1.311 | 0.005 | -0.007 |
| D | 0.20 | 0.430 | 0.022 | 0.705 | -0.002 | 0.008 | 0.305 | 0.028 | 0.580 | -0.004 | 0.006 |
| DD | 0.26 | 0.569 | -0.012 | 1.128 | -0.002 | 0.005 | 0.408 | -0.015 | 0.940 | -0.003 | 0.006 |
| DHI | 0.37 | 0.440 | 0.020 | 1.660 | 0.008 | 0.011 | 0.366 | 0.011 | 1.553 | 0.009 | 0.013 |
| DOV | 0.34 | 0.512 | 0.003 | 1.174 | 0.002 | 0.003 | 0.452 | 0.003 | 1.030 | 0.001 | 0.004 |
| FDX | 0.19 | 0.374 | 0.022 | 0.915 | 0.001 | 0.002 | 0.307 | 0.020 | 0.918 | 0.001 | 0.003 |
| FO | 0.40 | 0.420 | 0.018 | 0.738 | 0.000 | 0.004 | 0.306 | 0.017 | 0.742 | -0.001 | 0.005 |
| GM | 0.52 | 0.399 | -0.078 | 1.406 | 0.000 | 0.010 | 0.339 | -0.082 | 1.454 | 0.000 | 0.011 |
| HCBK | 0.34 | 0.415 | 0.083 | 0.648 | 0.002 | 0.006 | 0.344 | 0.079 | 0.711 | 0.003 | 0.007 |
| HCP | 0.37 | 0.515 | 0.044 | 0.733 | 0.004 | 0.006 | 0.398 | 0.030 | 1.002 | 0.006 | 0.012 |
| HPC | 0.34 | 0.330 | -0.011 | 1.279 | 0.004 | 0.009 | 0.242 | -0.007 | 1.097 | 0.002 | 0.008 |
| HSP | 0.44 | 0.307 | 0.019 | 0.825 | -0.002 | -0.003 | 0.215 | 0.015 | 0.648 | 0.000 | -0.003 |
| KFT | 0.53 | 0.406 | 0.005 | 0.457 | -0.003 | 0.002 | 0.213 | 0.002 | 0.478 | -0.001 | 0.002 |
| LMT | 0.28 | 0.206 | 0.035 | 0.611 | -0.002 | 0.005 | 0.146 | 0.036 | 0.569 | -0.002 | 0.004 |
| LOW | 0.22 | 0.455 | 0.049 | 1.118 | -0.001 | -0.001 | 0.363 | 0.041 | 1.125 | -0.001 | 0.003 |
| LSI | 0.38 | 0.442 | 0.046 | 1.682 | 0.011 | -0.008 | 0.355 | 0.050 | 1.656 | 0.009 | -0.008 |
| LXK | 0.46 | 0.261 | 0.019 | 0.823 | 0.004 | -0.005 | 0.197 | 0.022 | 0.862 | 0.004 | -0.006 |
| MDT | 0.55 | 0.295 | 0.014 | 0.647 | -0.003 | -0.002 | 0.213 | 0.012 | 0.615 | -0.003 | -0.001 |
| MI | 0.37 | 0.643 | -0.004 | 1.038 | -0.001 | 0.007 | 0.495 | -0.022 | 1.317 | 0.001 | 0.014 |
| MTW | 0.16 | 0.447 | 0.008 | 1.377 | 0.011 | 0.008 | 0.352 | 0.016 | 1.385 | 0.006 | 0.008 |
| MUR | 0.08 | 0.337 | 0.055 | 0.890 | 0.002 | 0.010 | 0.248 | 0.068 | 0.874 | -0.001 | 0.007 |
| MWW | 0.36 | 0.388 | 0.080 | 1.724 | 0.008 | -0.009 | 0.326 | 0.077 | 1.654 | 0.008 | -0.009 |
| NOV | 0.21 | 0.382 | 0.072 | 1.416 | 0.004 | 0.014 | 0.245 | 0.090 | 1.337 | 0.001 | 0.008 |
| NUE | 0.21 | 0.431 | 0.071 | 1.628 | 0.003 | 0.012 | 0.372 | 0.081 | 1.398 | 0.001 | 0.008 |
| ORCL | 0.15 | 0.461 | 0.116 | 1.262 | 0.000 | -0.013 | 0.412 | 0.117 | 1.301 | 0.000 | -0.013 |
| PAYX | 0.18 | 0.392 | 0.047 | 1.074 | -0.002 | -0.004 | 0.301 | 0.044 | 0.966 | -0.002 | -0.003 |
| PG | 0.27 | 0.299 | 0.035 | 0.533 | -0.005 | -0.001 | 0.175 | 0.028 | 0.463 | -0.005 | 0.002 |
| QCOM | 0.50 | 0.416 | 0.188 | 1.356 | 0.001 | -0.014 | 0.351 | 0.187 | 1.326 | 0.000 | -0.013 |
| S | 0.44 | 0.383 | -0.076 | 1.092 | -0.001 | 0.005 | 0.268 | -0.077 | 1.348 | -0.002 | 0.007 |
| S | 0.57 | 0.320 | 0.030 | 1.303 | -0.003 | 0.006 | 0.192 | 0.020 | 1.230 | -0.002 | 0.008 |
| SLE | 0.33 | 0.350 | -0.012 | 0.539 | -0.003 | 0.001 | 0.242 | -0.019 | 0.619 | -0.004 | 0.004 |
| SPG | 0.27 | 0.565 | 0.038 | 0.808 | 0.004 | 0.007 | 0.462 | 0.021 | 1.107 | 0.006 | 0.013 |
| SRE | 0.27 | 0.372 | 0.025 | 0.842 | -0.002 | 0.009 | 0.285 | 0.032 | 0.716 | -0.002 | 0.007 |
| SYMC | 0.56 | 0.297 | 0.124 | 1.085 | 0.005 | -0.007 | 0.241 | 0.122 | 1.100 | 0.005 | -0.007 |
| TDC | 0.36 | 0.610 | -0.021 | 0.981 | 0.003 | -0.001 | 0.432 | -0.027 | 0.901 | 0.000 | -0.003 |
| TE | 0.26 | 0.328 | -0.016 | 0.855 | -0.001 | 0.008 | 0.233 | -0.009 | 0.705 | -0.003 | 0.007 |
| TGT | 0.52 | 0.452 | 0.034 | 1.155 | -0.001 | 0.001 | 0.344 | 0.031 | 1.097 | -0.001 | 0.002 |
| TLAB | 0.24 | 0.391 | -0.019 | 1.780 | 0.004 | -0.002 | 0.302 | -0.008 | 1.403 | 0.003 | -0.007 |
| WFR | 0.31 | 0.276 | 0.148 | 1.815 | 0.018 | -0.003 | 0.186 | 0.156 | 1.734 | 0.013 | -0.003 |
| WLP | 0.31 | 0.392 | 0.048 | 0.601 | 0.000 | 0.001 | 0.193 | 0.051 | 0.674 | -0.002 | 0.001 |
| WLP | 0.34 | 0.123 | 0.071 | 0.675 | -0.001 | 0.005 | 0.085 | 0.072 | 0.641 | -0.002 | 0.006 |
| ZMH | 0.30 | 0.243 | 0.029 | 0.524 | 0.000 | -0.001 | 0.200 | 0.029 | 0.658 | 0.000 | -0.001 |
| Ave. | - | 0.382 | 0.034 | 1.041 | 0.002 | 0.002 | 0.284 | 0.034 | 1.025 | 0.001 | 0.002 |

| | | | | | Panel B - S&P MidCap 400 | | | | | |
|---|---|---|---|---|---|---|---|---|---|---|---|
| Ticker | p-value | $R^2_{KR}$ | $\hat{\alpha}_{KR}$ | $\hat{\beta}_{KR}$ | $\hat{s}_{KR}$ | $\hat{h}_{KR}$ | $R^2_{LR}$ | $\hat{\alpha}_{LR}$ | $\hat{\beta}_{LR}$ | $\hat{s}_{LR}$ | $\hat{h}_{LR}$ |
| AAI | 0.35 | 0.294 | 0.059 | 1.471 | 0.013 | 0.008 | 0.219 | 0.049 | 1.431 | 0.014 | 0.011 |
| AMG | 0.32 | 0.549 | 0.044 | 1.305 | 0.006 | 0.002 | 0.476 | 0.032 | 1.488 | 0.008 | 0.007 |
| ARG | 0.41 | 0.340 | 0.076 | 1.163 | 0.005 | 0.004 | 0.249 | 0.074 | 1.122 | 0.003 | 0.006 |
| AVCT | 0.35 | 0.437 | 0.046 | 1.275 | 0.012 | -0.006 | 0.309 | 0.062 | 1.378 | 0.011 | -0.009 |
| BRO | 0.34 | 0.399 | 0.060 | 0.765 | 0.004 | 0.003 | 0.218 | 0.061 | 0.703 | 0.002 | 0.004 |
| CLF | 0.32 | 0.424 | 0.082 | 1.460 | 0.009 | 0.012 | 0.313 | 0.089 | 1.537 | 0.005 | 0.012 |
| CMG | 0.36 | 0.428 | 0.144 | 1.235 | 0.009 | 0.001 | 0.276 | 0.137 | 1.025 | 0.010 | 0.001 |
| CPT | 0.43 | 0.537 | 0.018 | 0.675 | 0.005 | 0.006 | 0.441 | 0.001 | 0.949 | 0.007 | 0.012 |
| CR | 0.30 | 0.437 | -0.017 | 0.988 | 0.003 | 0.005 | 0.361 | -0.016 | 1.002 | 0.003 | 0.005 |
| CWTR | 0.19 | 0.196 | 0.072 | 1.240 | 0.012 | 0.010 | 0.174 | 0.069 | 1.431 | 0.013 | 0.011 |
| CXW | 0.21 | 0.173 | 0.015 | 0.802 | 0.008 | 0.004 | 0.070 | 0.019 | 0.753 | 0.007 | 0.003 |
| DCI | 0.37 | 0.395 | 0.048 | 0.914 | 0.005 | 0.002 | 0.335 | 0.045 | 0.900 | 0.004 | 0.004 |
| ELY | 0.44 | 0.277 | 0.018 | 1.146 | 0.007 | 0.003 | 0.233 | 0.015 | 1.067 | 0.006 | 0.005 |
| ENR | 0.36 | 0.267 | 0.059 | 0.731 | 0.002 | 0.002 | 0.189 | 0.042 | 0.762 | 0.003 | 0.005 |
| FMER | 0.42 | 0.586 | 0.004 | 1.053 | 0.003 | 0.006 | 0.449 | -0.008 | 1.186 | 0.004 | 0.011 |
| FNFG | 0.37 | 0.379 | 0.053 | 0.730 | 0.007 | 0.005 | 0.316 | 0.043 | 0.883 | 0.007 | 0.009 |
| FTO | 0.40 | 0.363 | 0.105 | 1.238 | 0.006 | 0.014 | 0.218 | 0.114 | 1.250 | 0.005 | 0.010 |
| HBI | 0.44 | 0.454 | 0.007 | 0.860 | 0.006 | 0.003 | 0.334 | 0.013 | 0.921 | 0.008 | 0.008 |
| HE | 0.33 | 0.396 | 0.005 | 0.578 | 0.001 | 0.007 | 0.215 | 0.016 | 0.439 | -0.001 | 0.003 |
| HMA | 0.12 | 0.182 | -0.042 | 0.677 | 0.002 | 0.004 | 0.103 | -0.059 | 0.758 | 0.004 | 0.009 |
| HNI | 0.49 | 0.458 | -0.013 | 0.941 | 0.006 | 0.006 | 0.316 | -0.019 | 1.074 | 0.005 | 0.009 |
| HRC | 0.51 | 0.201 | -0.018 | 0.490 | 0.001 | 0.002 | 0.138 | -0.018 | 0.545 | 0.000 | 0.003 |
| HRL | 0.41 | 0.225 | 0.033 | 0.591 | 0.000 | 0.002 | 0.102 | 0.033 | 0.411 | -0.001 | 0.002 |
| IDXX | 0.42 | 0.291 | 0.059 | 0.718 | 0.005 | -0.003 | 0.167 | 0.057 | 0.744 | 0.004 | -0.001 |
| IEX | 0.39 | 0.431 | 0.032 | 0.949 | 0.007 | 0.002 | 0.382 | 0.031 | 0.967 | 0.006 | 0.004 |
| IRF | 0.19 | 0.481 | 0.075 | 1.566 | 0.013 | -0.009 | 0.355 | 0.083 | 1.390 | 0.010 | -0.010 |
| JBHT | 0.21 | 0.311 | 0.065 | 1.331 | 0.009 | 0.007 | 0.244 | 0.074 | 1.094 | 0.006 | 0.004 |
| JBLU | 0.40 | 0.381 | 0.001 | 0.818 | 0.009 | -0.001 | 0.253 | -0.013 | 1.130 | 0.011 | 0.008 |
| KMT | 0.58 | 0.455 | 0.031 | 1.250 | 0.007 | 0.006 | 0.416 | 0.032 | 1.200 | 0.005 | 0.007 |
| LRCX | 0.13 | 0.461 | 0.146 | 1.722 | 0.010 | -0.014 | 0.393 | 0.141 | 1.616 | 0.009 | -0.012 |
| MEG | 0.45 | 0.419 | -0.103 | 0.864 | 0.009 | 0.006 | 0.241 | -0.131 | 1.309 | 0.013 | 0.015 |
| MRX | 0.38 | 0.261 | 0.000 | 0.761 | 0.005 | 0.001 | 0.178 | -0.001 | 0.833 | 0.006 | 0.001 |
| MTX | 0.38 | 0.469 | -0.013 | 1.038 | 0.007 | 0.006 | 0.358 | -0.008 | 0.933 | 0.004 | 0.005 |
| MVL | 0.36 | 0.131 | 0.109 | 0.812 | 0.008 | 0.004 | 0.106 | 0.112 | 0.854 | 0.008 | 0.003 |
| NHP | 0.43 | 0.447 | 0.030 | 0.890 | 0.006 | 0.007 | 0.393 | 0.023 | 0.978 | 0.007 | 0.009 |
| OII | 0.11 | 0.244 | 0.056 | 1.171 | 0.007 | 0.012 | 0.182 | 0.071 | 1.053 | 0.004 | 0.007 |
| ORLY | 0.21 | 0.331 | 0.055 | 0.947 | 0.004 | 0.000 | 0.209 | 0.055 | 0.918 | 0.003 | 0.002 |
| OSK | 0.39 | 0.328 | 0.039 | 1.018 | 0.006 | 0.004 | 0.208 | 0.033 | 1.042 | 0.006 | 0.006 |
| PSD | 0.41 | 0.259 | 0.009 | 0.588 | -0.001 | 0.006 | 0.176 | 0.012 | 0.450 | -0.001 | 0.005 |
| RS | 0.21 | 0.476 | 0.028 | 1.393 | 0.009 | 0.011 | 0.389 | 0.037 | 1.449 | 0.006 | 0.010 |
| RYL | 0.35 | 0.455 | 0.050 | 1.556 | 0.006 | 0.010 | 0.351 | 0.039 | 1.460 | 0.009 | 0.012 |
| SKS | 0.34 | 0.268 | -0.040 | 1.146 | 0.005 | 0.008 | 0.219 | -0.040 | 1.171 | 0.006 | 0.007 |
| SRCL | 0.39 | 0.179 | 0.116 | 0.733 | 0.004 | 0.001 | 0.114 | 0.112 | 0.686 | 0.004 | 0.003 |
| TECD | 0.30 | 0.299 | 0.002 | 1.125 | 0.005 | -0.002 | 0.221 | 0.006 | 0.974 | 0.004 | -0.004 |
| THG | 0.68 | 0.307 | -0.008 | 1.133 | 0.002 | 0.009 | 0.263 | -0.006 | 1.050 | 0.001 | 0.008 |
| UTHR | 0.45 | 0.271 | 0.136 | 0.784 | 0.006 | -0.001 | 0.113 | 0.136 | 0.773 | 0.010 | -0.004 |
| UTR | 0.53 | 0.474 | -0.019 | 0.985 | 0.002 | 0.008 | 0.396 | -0.019 | 1.054 | 0.001 | 0.009 |
| VARI | 0.35 | 0.409 | 0.083 | 1.181 | 0.010 | -0.003 | 0.278 | 0.080 | 1.134 | 0.010 | -0.002 |
| WBS | 0.32 | 0.554 | -0.021 | 0.934 | 0.002 | 0.006 | 0.445 | -0.028 | 1.142 | 0.002 | 0.009 |
| WOR | 0.48 | 0.383 | 0.014 | 1.441 | 0.004 | 0.010 | 0.342 | 0.017 | 1.282 | 0.004 | 0.008 |
| Ave. | - | 0.363 | 0.036 | 1.024 | 0.006 | 0.004 | 0.269 | 0.034 | 1.034 | 0.006 | 0.005 |

| Ticker | p-value | $R^2_{KR}$ | $\hat{\alpha}_{KR}$ | $\hat{\beta}_{KR}$ | $\hat{s}_{KR}$ | $\hat{h}_{KR}$ | $R^2_{LR}$ | $\hat{\alpha}_{LR}$ | $\hat{\beta}_{LR}$ | $\hat{s}_{LR}$ | $\hat{h}_{LR}$ |
|---|---|---|---|---|---|---|---|---|---|---|---|
| | | | | *Panel C - S&P SmallCap 600* | | | | | | | |
| ABM | 0.31 | 0.381 | 0.008 | 0.851 | 0.008 | 0.002 | 0.299 | -0.001 | 0.884 | 0.008 | 0.006 |
| ACLS | 0.30 | 0.564 | -0.042 | 1.824 | 0.020 | -0.007 | 0.309 | -0.044 | 1.800 | 0.017 | -0.005 |
| CASY | 0.12 | 0.433 | 0.026 | 1.104 | 0.008 | 0.003 | 0.230 | 0.028 | 0.895 | 0.006 | 0.003 |
| CCRN | 0.14 | 0.346 | -0.034 | 1.262 | 0.011 | 0.003 | 0.222 | -0.033 | 0.982 | 0.011 | 0.002 |
| DNEX | 0.50 | 0.386 | 0.014 | 0.888 | 0.010 | 0.000 | 0.232 | 0.020 | 0.851 | 0.007 | -0.001 |
| EE | 0.30 | 0.281 | 0.014 | 0.818 | 0.005 | 0.007 | 0.233 | 0.019 | 0.716 | 0.004 | 0.005 |
| GFF | 0.42 | 0.306 | 0.003 | 0.953 | 0.010 | 0.004 | 0.243 | -0.002 | 0.963 | 0.009 | 0.007 |
| GTIV | 0.36 | 0.328 | 0.140 | 0.718 | 0.008 | 0.002 | 0.166 | 0.138 | 0.707 | 0.008 | 0.002 |
| HMSY | 0.56 | 0.082 | 0.106 | 0.553 | 0.004 | 0.001 | 0.033 | 0.104 | 0.509 | 0.005 | 0.002 |
| HOMB | 0.35 | 0.539 | 0.127 | 0.924 | 0.011 | 0.002 | 0.397 | 0.123 | 0.796 | 0.013 | 0.005 |
| HTLD | 0.21 | 0.333 | 0.049 | 1.098 | 0.010 | 0.005 | 0.247 | 0.052 | 0.913 | 0.009 | 0.004 |
| INT | 0.35 | 0.299 | 0.088 | 0.970 | 0.008 | 0.008 | 0.226 | 0.088 | 1.139 | 0.006 | 0.010 |
| ISYS | 0.25 | 0.252 | 0.040 | 0.910 | 0.007 | 0.004 | 0.131 | 0.042 | 0.885 | 0.009 | 0.002 |
| KNOT | 0.25 | 0.130 | -0.065 | 1.090 | 0.006 | 0.006 | 0.043 | -0.071 | 0.849 | 0.008 | 0.005 |
| MANT | 0.39 | 0.175 | 0.085 | 0.718 | 0.010 | -0.001 | 0.125 | 0.084 | 0.721 | 0.009 | 0.001 |
| MFB | 0.36 | 0.477 | -0.004 | 1.250 | 0.006 | 0.009 | 0.354 | -0.001 | 1.157 | 0.013 | 0.005 |
| MNT | 0.36 | 0.147 | 0.073 | 0.728 | 0.005 | 0.001 | 0.042 | 0.081 | 0.460 | 0.003 | -0.002 |
| MNT | 0.29 | 0.015 | -0.004 | 0.202 | 0.000 | 0.003 | 0.025 | -0.007 | 0.423 | 0.000 | 0.005 |
| MOH | 0.48 | 0.141 | 0.022 | 0.442 | 0.006 | -0.002 | 0.107 | 0.013 | 0.654 | 0.005 | 0.002 |
| MTH | 0.24 | 0.383 | 0.079 | 1.657 | 0.012 | 0.015 | 0.303 | 0.069 | 1.717 | 0.014 | 0.017 |
| NILE | 0.42 | 0.267 | 0.056 | 1.092 | 0.007 | -0.005 | 0.183 | 0.037 | 0.956 | 0.013 | 0.003 |
| NOVN | 0.60 | 0.187 | 0.098 | 0.907 | 0.013 | -0.003 | 0.106 | 0.101 | 0.923 | 0.010 | -0.002 |
| NPK | 0.42 | 0.410 | 0.027 | 0.589 | 0.007 | 0.003 | 0.341 | 0.022 | 0.734 | 0.006 | 0.006 |
| NSIT | 0.41 | 0.297 | 0.013 | 1.254 | 0.011 | -0.001 | 0.207 | 0.012 | 1.311 | 0.011 | 0.000 |
| NTRI | 0.22 | 0.074 | 0.019 | 0.780 | 0.009 | 0.004 | 0.047 | 0.020 | 0.767 | 0.009 | 0.004 |
| NWK | 0.46 | 0.256 | 0.005 | 1.355 | 0.015 | 0.003 | 0.135 | 0.012 | 1.116 | 0.012 | 0.002 |
| ONB | 0.33 | 0.472 | -0.020 | 0.820 | 0.007 | 0.007 | 0.381 | -0.028 | 0.886 | 0.007 | 0.010 |
| PBY | 0.61 | 0.259 | -0.032 | 1.250 | 0.009 | 0.011 | 0.226 | -0.027 | 1.281 | 0.009 | 0.010 |
| PEI | 0.51 | 0.585 | -0.018 | 0.775 | 0.008 | 0.008 | 0.412 | -0.027 | 1.191 | 0.007 | 0.014 |
| PENX | 0.30 | 0.135 | 0.021 | 0.730 | 0.003 | 0.004 | 0.092 | 0.013 | 0.804 | 0.005 | 0.006 |
| PRFT | 0.20 | 0.069 | 0.202 | 0.648 | 0.006 | -0.002 | 0.037 | 0.186 | 1.027 | 0.008 | 0.004 |
| PRGS | 0.23 | 0.341 | 0.020 | 1.012 | 0.011 | -0.001 | 0.238 | 0.020 | 0.992 | 0.010 | 0.000 |
| PVTB | 0.28 | 0.314 | 0.067 | 0.666 | 0.008 | 0.004 | 0.178 | 0.054 | 0.744 | 0.008 | 0.008 |
| RADS | 0.34 | 0.232 | 0.063 | 1.235 | 0.012 | 0.001 | 0.169 | 0.066 | 1.395 | 0.010 | 0.001 |
| RBN | 0.15 | 0.289 | 0.022 | 0.966 | 0.008 | 0.004 | 0.215 | 0.021 | 0.955 | 0.006 | 0.006 |
| SAFM | 0.55 | 0.250 | 0.071 | 0.699 | 0.006 | 0.003 | 0.112 | 0.065 | 0.723 | 0.006 | 0.005 |
| SKT | 0.41 | 0.436 | 0.061 | 0.753 | 0.005 | 0.007 | 0.374 | 0.054 | 0.841 | 0.006 | 0.009 |
| SLXP | 0.38 | 0.262 | 0.086 | 1.035 | 0.012 | -0.005 | 0.103 | 0.073 | 0.909 | 0.011 | 0.000 |
| SNS | 0.40 | 0.344 | -0.057 | 1.035 | 0.012 | 0.008 | 0.207 | -0.050 | 0.904 | 0.010 | 0.006 |
| SNS | 0.66 | 0.082 | 0.257 | 0.561 | -0.002 | 0.000 | 0.070 | 0.257 | 0.682 | -0.002 | 0.003 |
| SSS | 0.28 | 0.508 | 0.024 | 0.708 | 0.007 | 0.006 | 0.437 | 0.016 | 0.905 | 0.007 | 0.009 |
| SUP | 0.30 | 0.414 | -0.039 | 0.918 | 0.005 | 0.006 | 0.284 | -0.037 | 0.902 | 0.004 | 0.006 |
| SYMM | 0.25 | 0.366 | 0.051 | 1.282 | 0.019 | 0.000 | 0.225 | 0.048 | 1.354 | 0.019 | 0.001 |
| TSFG | 0.34 | 0.505 | -0.049 | 1.061 | 0.008 | 0.008 | 0.345 | -0.068 | 1.420 | 0.010 | 0.015 |
| TXRH | 0.30 | 0.451 | 0.003 | 1.103 | 0.009 | 0.001 | 0.262 | -0.004 | 0.847 | 0.012 | 0.003 |
| UEIC | 0.52 | 0.264 | 0.071 | 1.040 | 0.010 | 0.001 | 0.202 | 0.064 | 1.063 | 0.012 | 0.003 |
| UNS | 0.44 | 0.293 | 0.026 | 0.641 | 0.004 | 0.006 | 0.233 | 0.028 | 0.671 | 0.003 | 0.006 |
| VECO | 0.29 | 0.434 | -0.014 | 1.612 | 0.015 | -0.007 | 0.308 | -0.011 | 1.474 | 0.013 | -0.007 |
| VSEA | 0.25 | 0.444 | 0.153 | 1.585 | 0.012 | -0.015 | 0.354 | 0.145 | 1.496 | 0.012 | -0.012 |
| ZEP | 0.42 | 0.561 | 0.375 | 1.224 | 0.007 | -0.004 | 0.451 | 0.382 | 1.268 | 0.011 | -0.008 |
| Ave. | - | 0.316 | 0.046 | 0.966 | 0.009 | 0.002 | 0.218 | 0.043 | 0.971 | 0.009 | 0.004 |

In Panel A, B and C can be seen the estimated parameters and statistics of the Fama-French three-factor model of 50-50 randomly chosen companies from the S&P 500, S&P MidCap 400 and the S&P SmallCap 600 universe. We estimate two models: 1, the kernel regression, the non-parametric Fama-French model is estimated. We select the bandwidth matrix optimally by Cross Validation (CV) (We use the Gaussian kernel and the Nadaraya (1964) and Watson (1964) weighting function in the kernel regression); 2, the standard Fama-French regression. The market return is the Center for Research in Security Prices (CRSP) value weighted index return; the risk-free rate is the return of the one-month Treasury bill. In the first column can be seen the tickers of each stock in the database, in column 2 the *p*-value of the linearity test, in column 3 the $R^2$s, then alphas and estimated factor loadings of the kernel and the linear regressions.

Figure 1 Characteristic "curves"

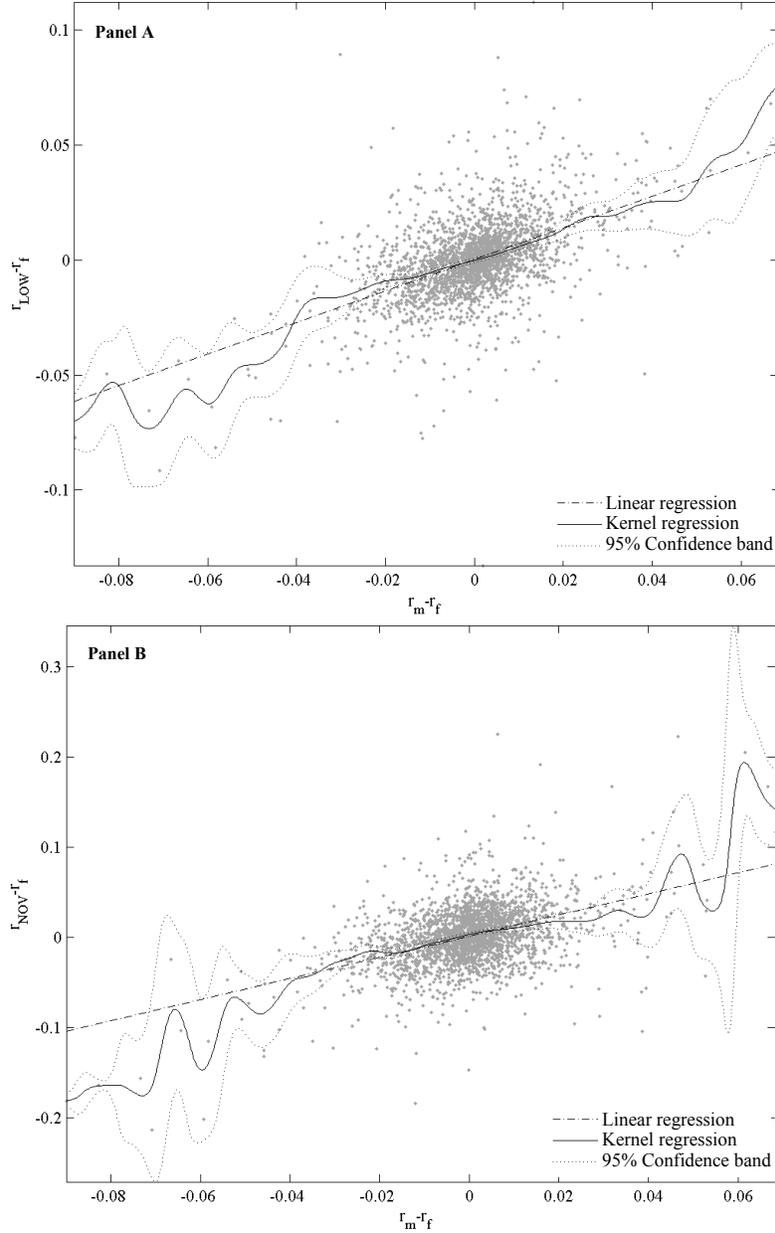

*The characteristic curve of the Lowe's Companies Inc. (Panel A) and the National Oilwell Varco Inc. (Panel B)*
*Notes:* We estimate the characteristic curve of both companies by two different methods: 1, by the kernel regression, the non-parametric CAPM is estimated (bold curve), that is $r_j - r_f = \hat{m}_h(r_m - r_f) + \hat{\eta}_j$ where $\hat{\eta}_j$ are the residuals, $r_j$-$r_f$ and $r_M$-$r_f$ are the risk premium of the stock and the market respectively, $h$ is an optimally selected bandwidth by Cross Validation (CV) (We use the Gaussian kernel and the Nadaraya (1964) and Watson (1964) weighting function in the kernel regression); 2, by linear regression (dashed line), that is $r_j - r_f = \hat{\alpha}_j + \hat{\beta}_j(r_m - r_f) + \hat{\varepsilon}_j$ is estimated where $\hat{\varepsilon}_j$ is the residual series. The market return is the Center for Research in Security Prices (CRSP) value weighted index return; the risk-free rate is the return of the one-month Treasury bill. The dotted lines represent the 95% confidence bands (see Härdle, 2004 for the calculation).

Figure 2 Semi-parametric derivative estimation

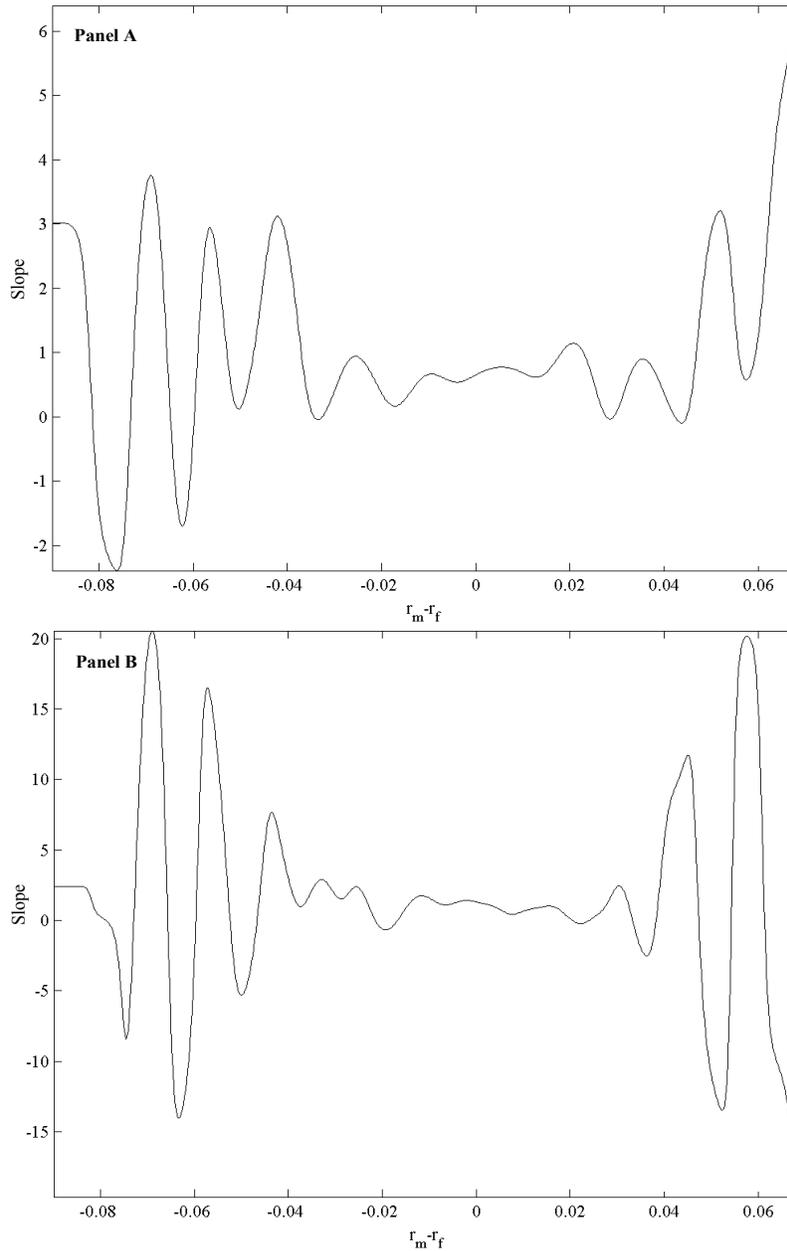

The figure shows the estimated derivative in the function of the risk premium. The semi-parametric derivative estimation is obtained by the weighted least squares estimator using the Nadaraya-Watson kernel weights. The characteristic curves of the Lowe's Companies Inc. (Panel A) and the National Oilwell Varco Inc. (Panel B) (randomly chosen stocks) are estimated by the kernel regressions. The market return is the Center for Research in Security Prices (CRSP) value weighted index return; the risk-free rate is the return of the one-month Treasury bill. The model, the non-parametric CAPM, is in the form $r_j - r_f = \hat{m}_h(r_m - r_f) + \hat{\eta}_j$ where $\hat{\eta}_j$ are the residuals, $r_j$-$r_f$ and $r_M$-$r_f$ are the risk premium of the stock and the market respectively, $h$ is an optimally selected bandwidth by Cross Validation (CV). We use the Gaussian kernel and the Nadaraya (1964) and Watson (1964) weighting function in the kernel regression.

Figure 3 Semi-parametric alpha estimation

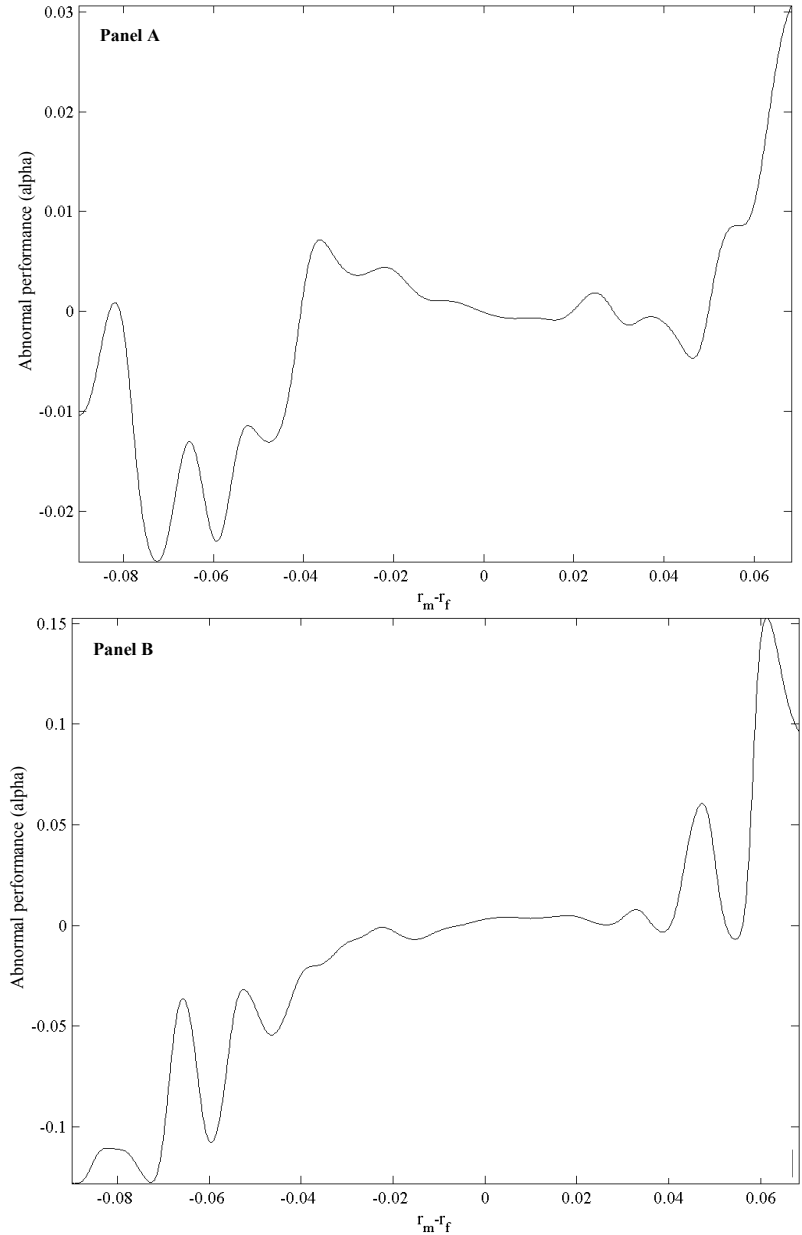

The figure shows the estimated semi-parametric alpha (the abnormal return) in the function of the risk premium. The semi-parametric derivative estimation is obtained by the weighted least squares estimator using the Nadaraya-Watson kernel weights. The characteristic curves of the Lowe's Companies Inc. (Panel A) and the National Oilwell Varco Inc. (Panel B) (randomly selected stocks) are estimated by the kernel regressions. The market return is the Center for Research in Security Prices (CRSP) value weighted index return; the risk-free rate is the return of the one-month Treasury bill. The model, the non-parametric CAPM is in the form $r_j - r_f = \hat{m}_h(r_m - r_f) + \hat{\eta}_j$ where $\hat{\eta}_j$ are the residuals, $r_j$-$r_f$ and $r_M$-$r_f$ are the risk premium of the stock and the market respectively, $h$ is an optimally selected bandwidth by Cross Validation (CV). We use the Gaussian kernel and the Nadaraya (1964) and Watson (1964) weighting function in the kernel regression.

Figure 4 Security market lines for the S&P 500, the S&P MidCap 400, the S&P SmallCap 600 and for all the stocks

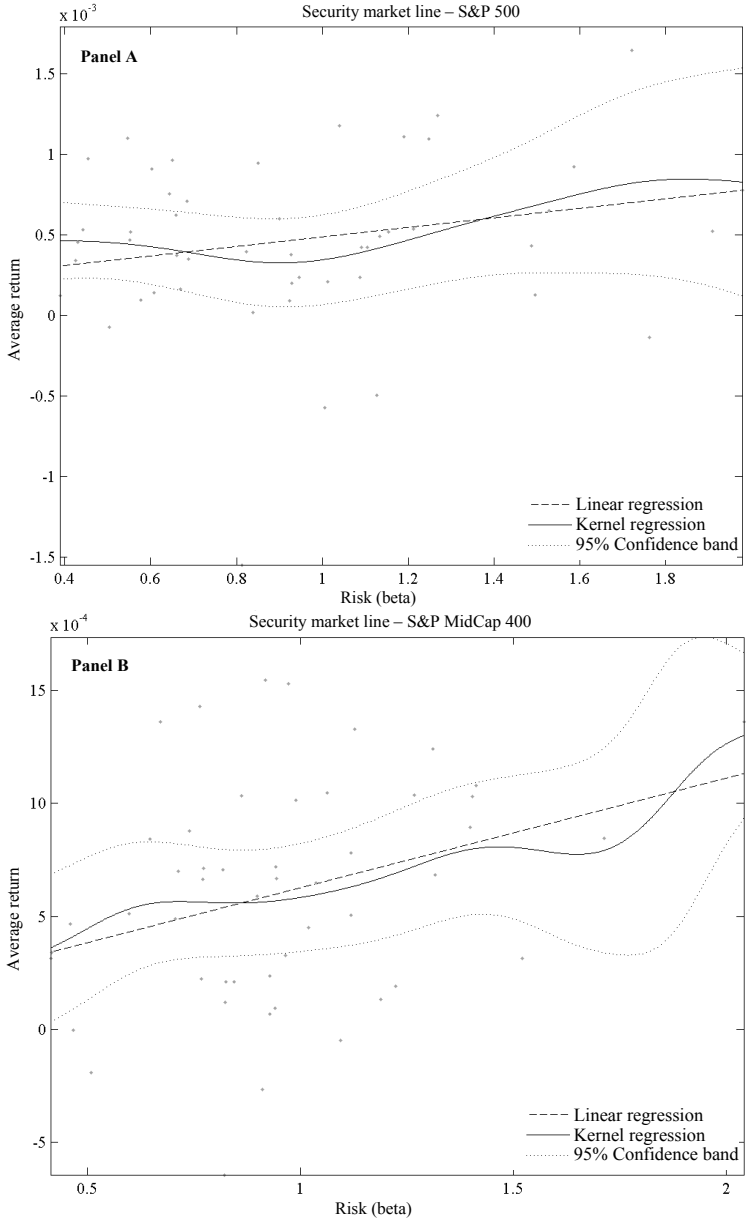

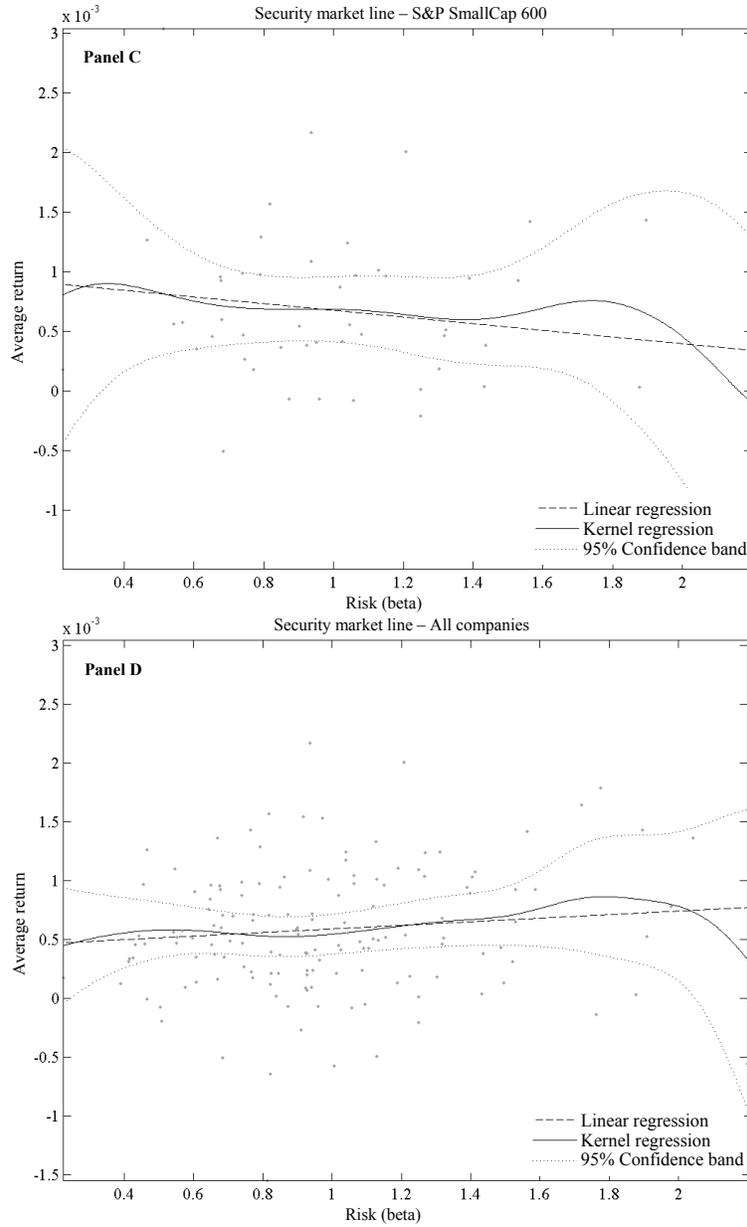

We estimate security market lines for the S&P 500, S&P MidCap 400, S&P SmallCap 600 stocks and for all the stocks in our database in Panel A, B, C, D respectively. We estimate the security market lines by two methods: 1, by the kernel regressions (bold curves) in the form $\bar{r}_j = \hat{m}_h(\hat{\beta}_j^*) + \hat{\lambda}_j$, where $\bar{r}_j$ is the average return of stock $j$ in the investigated period, $\hat{\beta}_j^*$ is the semi-parametric beta of stock $j$, $\hat{\lambda}_j$ is the residual series and $h$ is an optimally selected bandwidth by Cross Validation (CV) (We use the Gaussian kernel and the Nadaraya (1964) and Watson (1964) weighting function in the kernel regression.); 2, by linear regression (dashed lines) in the form $\bar{r}_j = \hat{\alpha}'_j + \hat{\beta}'_j \hat{\beta}_j^* + \hat{\kappa}_j$, where $\hat{\kappa}_j$ are the residuals. The market return is the Center for Research in Security Prices (CRSP) value weighted index return; the risk-free rate is the return of the one-month Treasury bill. The dotted curves represent the 95% confidence bands (see Härdle, 2004 for the calculation).